\documentclass[aps,prc,groupedaddress,twocolumn,nofootinbib]{revtex4-2}

\usepackage{bm}
\usepackage{graphicx}
\usepackage{ascmac}
\usepackage{amsmath,amssymb}
\usepackage{cancel}
\usepackage{braket}
\renewcommand{\Re}{\operatorname{Re}}
\renewcommand{\Im}{\operatorname{Im}}

\begin{document}


\title{Compositeness of near-threshold eigenstates with Coulomb plus short-range interactions}


\author{Tomona Kinugawa}
\email[]{tomona.kinugawa@riken.jp}
\affiliation{Nishina Center for Accelerator-Based Science, RIKEN, Wako 351-0198, Japan}
\author{Tetsuo Hyodo}
\email[]{hyodo@rcnp.osaka-u.ac.jp}
\affiliation{Research Center for Nuclear Physics, The University of Osaka, Ibaraki, Osaka 567-0047, Japan}

\date{\today}

\begin{abstract}
We investigate the internal structure of near-threshold $s$-wave eigenstates in a two-body system with Coulomb plus short-range interactions. Using a nonrelativistic effective field theory, we derive the expression for the compositeness in terms of the energy derivative of the self-energy, which is applicable to the present system with the non-separable Coulomb interaction. For near-threshold states, the compositeness can be written solely in terms of the Coulomb scattering length, the Coulomb effective range, and the Bohr radius, providing the weak-binding relation in the presence of the Coulomb interaction. We numerically study the pole trajectories and the compositeness and find that the Coulomb interaction qualitatively modifies the threshold behavior of the poles and the internal structure of the eigenstates. We show that when the Coulomb interaction is relatively strong, the enhancement of the compositeness near the threshold is absent, in contrast to purely short-range interactions. On the other hand, for a weak Coulomb interaction, a remnant of short-range universality survives, and near-threshold bound states tend to be composite dominant. Furthermore, even resonances are dominated by the composite component in the presence of the Coulomb interaction, owing to their continuous connection to the bound-state regime. We apply the formalism to realistic systems with near-threshold eigenstates, including exotic hadrons and nuclei.

\end{abstract}


\maketitle


\section{Introduction}
\label{sec:intro}

Understanding the internal structure of exotic hadrons is one of the central goals in hadron physics. In recent years, numerous exotic hadron candidates have been discovered~\cite{ParticleDataGroup:2024cfk}, which cannot be classified within the conventional quark model of mesons ($q\bar q$) and baryons ($qqq$). These states are expected to have more complex structures, such as compact multiquark configurations and hadronic molecules. The latter are weakly bound systems of two hadrons that retain their identities as color-singlet degrees of freedom~\cite{Hosaka:2016pey,Guo:2017jvc,Brambilla:2019esw,Hanhart:2025bun,Hosaka:2025gcl}. Notably, many exotic hadrons are observed near two-body scattering thresholds, indicating that the study of near-threshold states provides an essential clue to unveiling the underlying structure of exotic hadrons.

To quantify the internal structure of such states, the compositeness $X$ has been introduced as a measure of the molecular component in the wavefunction~\cite{Weinberg:1965zz,Baru:2003qq,Hyodo:2011qc,Aceti:2012dd,Hyodo:2013nka,Oller:2017alp,vanKolck:2022lqz,Kinugawa:2024crb}. For a bound state, $X$ is defined as the overlap of the bound-state wavefunction with the two-body scattering states. The compositeness $X$ represents the probability of finding the molecular component, while the remaining fraction $Z = 1 - X$, called the elementarity, accounts for non-molecular components such as compact multiquark configurations. In this way, the compositeness provides a useful framework for characterizing the structure of near-threshold states, and concepts analogous to the compositeness have been applied not only in hadron physics but also in nuclear and atomic physics~\cite{Duine:2003zza,Duine:2003zz,Braaten:2003sw,Partridge:2005zz,Schmidt:2011zu,Kinugawa:2022fzn,nwab226,Naidon:2024bdy}. 

When the interaction between the constituents is of short range, near-threshold systems exhibit universal properties governed by the large scattering length~\cite{Braaten:2004rn,Naidon:2016dpf}. In this regime, the detailed microscopic dynamics becomes irrelevant, and the nature of the states can be understood in a model-independent manner. For instance, as the binding energy approaches zero, the bound state becomes purely composite with $X \to 1$~\cite{Hyodo:2014bda}. More generally, shallow bound states tend to be composite dominant, whereas near-threshold resonances slightly above the threshold tend to be non-composite dominant~\cite{Hyodo:2013iga,Matuschek:2020gqe,Hanhart:2022qxq,Kinugawa:2023fbf,Kinugawa:2024kwb}. This low-energy universality provides a unified understanding of the structure of near-threshold states in systems with short-range interactions. For example, it is expected that the empirical observation in hadron physics that hadronic molecular states are found near thresholds~\cite{Hosaka:2016pey,Guo:2017jvc,Brambilla:2019esw,Hanhart:2025bun} and the so-called threshold rule in nuclear physics, where $\alpha$-cluster states emerge near thresholds~\cite{Ikeda:1968fry}, originate from the same underlying mechanism~\cite{Hosaka:2025gcl}.

In realistic systems, however, additional interactions may be present. In particular, when the constituents are electrically charged, the Coulomb interaction acts together with the short-range force. Although the electromagnetic interaction is typically much weaker than the strong interaction, its effects become non-negligible in near-threshold states with small binding energies. In fact, since the $\alpha$ clusters mentioned above have charge $+2$, it is not necessarily obvious whether the origin of the threshold rule in nuclear systems can be understood solely on the basis of the universality of short-range interactions. In addition, the large masses and charges of nuclei can make the Bohr radius as small as the interaction range, so that the effect of the Coulomb interaction on shallow bound states is expected to become significant. Furthermore, several exotic hadrons, such as $X(3872)$ and $T_{cc}$, have binding energies of the order of tens to hundreds of keV~\cite{ParticleDataGroup:2024cfk,LHCb:2021vvq,LHCb:2021auc}, which are comparable to the characteristic Coulomb scale. In such cases, even a weak Coulomb interaction can significantly affect the properties of the states.

In extreme cases, the Coulomb interaction can qualitatively change the nature of near-threshold states. For instance, the ${}^{8}\mathrm{Be}$ nucleus appears as a near-threshold resonance in the $\alpha\alpha$ system, while it would form a bound state if the Coulomb repulsion were switched off~\cite{Braaten:2004rn,Higa:2008dn}. Similarly, lattice QCD studies indicate that the $\Omega_{ccc}^{++}\Omega_{ccc}^{++}$ system forms a bound state with only the strong interaction but the state becomes unbound when the Coulomb repulsion is included~\cite{Lyu:2021qsh}. On the other hand, the Coulomb attraction can also assist binding, as in the $\Xi^{-}$-$\alpha$ system, where a Coulomb-assisted bound state emerges even when the strong interaction alone is insufficient~\cite{Hiyama:2022jqh,Kamiya:2024diw}. In such cases, the internal structure of the eigenstate may also be significantly modified by the Coulomb interaction.

The presence of the long-range Coulomb interaction modifies the low-energy behavior of the scattering amplitude, and the standard low-energy universality derived for short-range interactions no longer holds~\cite{Bethe:1949yr,Domcke:1983zz,Kong:1998sx,Kong:1999sf,Ando:2007fh,Higa:2008dn,J.Phys.Chem.123.82,Schmickler:2019dcy,Mochizuki:2024dbf,Albaladejo:2025kuv}. Therefore, it is an important theoretical problem to clarify how the properties of near-threshold states, including their internal structure and compositeness, are affected by the interplay between the Coulomb and short-range interactions. In particular, the threshold rule empirically found in nuclear structure studies cannot be rigorously justified by the universality derived for short-range interactions alone, and it is necessary to establish its theoretical foundation by taking into account the Coulomb interaction between the constituent particles.

In this work, we investigate near-threshold $s$-wave eigenstates in a two-body system with Coulomb plus short-range interactions. We analyze the scattering amplitude and its analytic structure, with particular emphasis on the pole structure in the complex momentum plane. Furthermore, we evaluate the compositeness of the eigenstates and discuss how the internal structure is modified by the Coulomb interaction. Through these analyses, we aim to provide a systematic understanding of near-threshold phenomena beyond the framework of short-range interactions.

This paper is organized as follows. In Sec.~\ref{sec:EFT}, we introduce the theoretical framework for systems with Coulomb plus short-range interactions. Based on the scattering amplitude, we derive the condition for the eigenstates and present an explicit expression for the compositeness. In Secs.~\ref{sec:negative-re} and \ref{sec:positive-re}, we numerically evaluate the compositeness of the eigenstates for systems with negative and positive effective ranges, respectively, and discuss their general properties. Both attractive and repulsive Coulomb interactions are considered. In Sec.~\ref{sec:apply}, we apply the framework to realistic systems and discuss the internal structure of the states. Finally, a summary is given in Sec.~\ref{sec:sum}.


\section{Effective field theory for systems with Coulomb plus short-range interactions}
\label{sec:EFT}

In this section, we introduce a nonrelativistic effective field theory (EFT) for the near-threshold scattering with the Coulomb plus short-range interactions~\cite{Kong:1998sx,Kong:1999sf,Ando:2007fh,Higa:2008dn}. In Sec.~\ref{subsec:Hamiltonian}, we outline the treatment of the Coulomb interaction in the framework of the EFT. The scattering amplitude is computed in Sec.~\ref{subsec:amplitude} and the pole condition is derived in Sec.~\ref{subsec:pole-condition}. As a quantitative measure to investigate the internal structure of near-threshold states, we introduce the compositeness in Sec.~\ref{subsec:compositeness}.

\subsection{Hamiltonian and Green function}
\label{subsec:Hamiltonian}

\begin{figure}[tbp]
\centering
\includegraphics[width=0.25\textwidth]{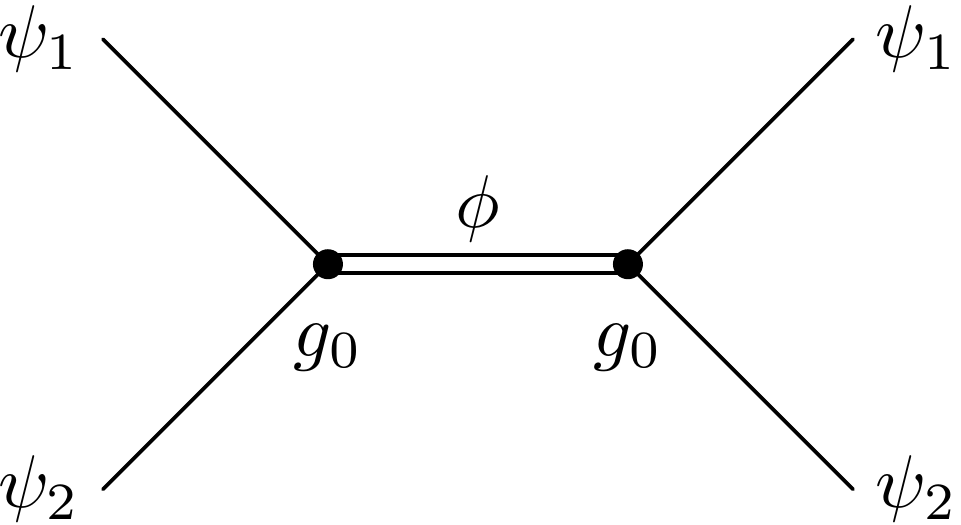}
\caption{Diagram of $\psi_{1}\psi_{2}$ scattering through the $s$-channel exchange of $\phi$.}
\label{fig:diagram}
\end{figure}

Here we construct a model to describe systems with the Coulomb plus short-range interactions, starting from the EFT Hamiltonian with only the short-range interaction. The contribution of the Coulomb interaction will be incorporated later. We introduce a free Hamiltonian $H_{0}$ and short-range interaction Hamiltonian $H_{S}$ in which the $s$-wave scattering states of $\psi_{1}$ and $\psi_{2}$ couple to a discrete state $\phi$~\cite{Braaten:2007nq,Kamiya:2015aea,Kamiya:2016oao,Kinugawa:2022fzn,Kinugawa:2023fbf}:
\begin{align}
H_{0} 
&= \int d^{3}r\ \left[\frac{1}{2m_{1}}{\nabla}\psi_{1}^{\dag}\cdot {\nabla}\psi_{1}+\frac{1}{2m_{2}}{\nabla}\psi_{2}^{\dag}\cdot {\nabla}\psi_{2} \right. \nonumber \\
& \quad + \left. \sigma\left(\frac{1}{2M}{\nabla}\phi^{\dag}\cdot {\nabla}\phi+\nu_{0}\phi^{\dag}\phi\right)\right], 
\label{eq:H-free-SR}\\
H_{S}
&= \int d^{3}r\ \left[g_{0}(\phi^{\dag}\psi_{1}\psi_{2}+\psi^{\dag}_{1}\psi^{\dag}_{2}\phi)\right].
\label{eq:H-int-SR}
\end{align}
Here $m_{1}$, $m_{2}$, and $M$ are the masses of $\psi_{1}$, $\psi_{2}$, and $\phi$, respectively. $\nu_{0}$ represents the energy of $\phi$ measured from the scattering threshold of $\psi_{1}$ and $\psi_{2}$. $g_{0}$ is a bare coupling constant between $\psi_{1,2}$ and $\phi$. $\sigma = \pm 1$ is introduced to describe systems with either a positive or a negative effective range~\cite{Kaplan:1996nv,Braaten:2007nq}. When $\sigma = +1$, $\phi$ represents the usual bare field with a positive norm, and the effective range is necessarily negative. On the other hand, for $\sigma = -1$, $\phi$ denotes the ghost field with a negative norm, which allows one to obtain a positive effective range. From the three-point interactions in $H_{S}$, the short-range $\psi_{1}\psi_{2}$ interaction occurs through the $s$-channel exchange of $\phi$ (see Fig.~\ref{fig:diagram}). Adopting the three-point interactions in $H_{S}$ is useful for introducing the compositeness in later sections~\cite{Kinugawa:2024crb}.

Now we assume that $\psi_{i}$ carries an electric charge $Z_{i}e$. The Coulomb potential $V_{C}$ between $\psi_{1}$ at $\bm{r}_{1}$ and $\psi_{2}$ at $\bm{r}_{2}$ is given by
\begin{align}
V_{C}(|\bm{r}_{1} - \bm{r}_{2}|) &= \frac{Z_{1}Z_{2}\alpha}{|\bm{r}_{1} - \bm{r}_{2}|}.
\label{eq:HC}
\end{align}
Here $\alpha = e^{2}/(4\pi) \sim 1/137$ is the fine-structure constant. This Coulomb potential in Eq.~\eqref{eq:HC} can be obtained from the following EFT Hamiltonian~\cite{Braaten:2007nq}:
\begin{align}
H_{C} &= \int d^{3}r_{1} d^{3}r_{2} \psi_{1}^{\dag}(\bm{r}_{1})\psi_{1}(\bm{r}_{1}) \psi_{2}^{\dag}(\bm{r}_{2})\psi_{2}(\bm{r}_{2}) V_{C}(|\bm{r}_{1} - \bm{r}_{2}|).
\label{eq:H-int-C}
\end{align}
Thus, the EFT defined by $H_{0} + H_{S} + H_{C}$ describes systems with Coulomb and short-range interactions.

In the EFT for short-range interactions, the two-body problem can be solved by the Lippmann-Schwinger (LS) equation to obtain the scattering amplitude. In fact, since the $\psi_{1}\psi_{2}$ potential derived from $H_{S}$ in Eq.~\eqref{eq:H-int-SR} is separable, the scattering amplitude can be obtained analytically~\cite{Braaten:2007nq}. However, the Coulomb interaction $H_{C}$ is not separable, and hence the LS equation for $H_{S}+H_{C}$ takes the form of an integral equation. Even in this case, the scattering amplitude can be analytically obtained by introducing the Coulomb correction in the Green function. In practice, in the LS equation with the potential derived from $H_{S}$, we replace the free Green function $G_{0}$ of the short-range case
\begin{align}
G_{0} &= \frac{1}{E - H_{0} + i0^{+}},
\label{eq:G0}
\end{align}
with the Coulomb Green function~\cite{Kong:1998sx,Kong:1999sf,Ando:2007fh,Higa:2008dn}
\begin{align}
G_{C} &= \frac{1}{E - (H_{0} + H_{C}) + i0^{+}}.
\label{eq:GC}
\end{align} 

\subsection{Scattering amplitude}
\label{subsec:amplitude}

To study the properties of the near-threshold eigenstates, we need to determine the complex eigenmomenta associated with these states. For this purpose, let us calculate the $s$-wave scattering amplitude $f(k)$ for the momentum $k$ whose pole positions correspond to the discrete eigenmomenta of the system~\cite{Taylor}. We follow the derivation of $f(k)$ in Ref.~\cite{Higa:2008dn}, and extend it to include both repulsive and attractive Coulomb interactions.

To facilitate the calculation, we compute the scattering amplitude by separating the pure-Coulomb contribution $f_{C}(k)$ from the Coulomb-distorted short-range contribution $f_{CS}(k)$~\cite{Taylor,Higa:2008dn}:
\begin{align}
f(k) &= f_{C}(k) + f_{CS}(k).
\label{eq:decomposition}
\end{align}
In the following, we show the explicit forms of $f_{C}(k)$ and $f_{CS}(k)$ separately.

\subsubsection{Pure-Coulomb scattering amplitude}
We first review the pure-Coulomb amplitude $f_{C}(k)$. For convenience, we introduce the Sommerfeld parameter $\eta$, which characterizes the contribution of the Coulomb interaction to the scattering process at momentum $k$:
\begin{align}
\eta(k) &= \frac{k_{C}}{k}.
\label{eq:eta}
\end{align}
Here, the parameter $k_{C}$ and the Bohr radius $a_{B}$ are defined as
\begin{align}
k_{C} &= \mu \alpha Z_1 Z_2,\\
a_B &= \frac{1}{|k_{C}|} = \frac{1}{\mu \alpha |Z_1 Z_2|},
\label{eq:aB}
\end{align}
with the reduced mass $\mu = 1/(1/m_{1} + 1/m_{2})$ and the product of the electric charges $Z_{1}Z_{2}$. Note that $a_{B}$ is defined not only for attractive but also for repulsive Coulomb interactions. In this case, $a_{B}$ can be regarded as a typical length scale for characterizing the strength of the Coulomb interaction. While $a_{B}$ in Eq.~\eqref{eq:aB} is always positive, $k_{C}$ and $\eta(k)$ are defined to be negative (positive) for attractive (repulsive) Coulomb interactions. 
In Table~\ref{tab:quantities-sum}, we summarize the signs of the quantities introduced here.

\begin{table}
 \caption{Summary of the signs of the quantities in the present formulation for attractive and repulsive Coulomb interactions.\label{tab:quantities-sum}}
 \begin{ruledtabular}
  \begin{tabular}{ccccccc}
    Coulomb interaction & $Z_{1}Z_{2}$ & $a_{B}$ & $k_{C}$ & $\eta(k)$ \\ \hline 
    Attractive & $-$ & $+$ & $-$ & $-$ \\
    Repulsive & $+$ & $+$ & $+$ & $+$ \\
  \end{tabular}
  \end{ruledtabular}
\end{table}

In general, the pure Coulomb scattering amplitude for the $l$-th partial wave is given by~\cite{Taylor}
\begin{align}
f_{C,l}(k) &= \frac{e^{2i\sigma_{l}(k)}-1}{2ik},
\label{eq:f-l}
\end{align}
with the Coulomb phase shift 
\begin{align}
\sigma_{l}(k) &= \arg\Gamma[l + 1 + i\eta(k)],
\label{eq:sigma-l}
\end{align}
for $k>0$. Then the pure-Coulomb scattering amplitude $f_{C}(k,\theta)$ at the scattering angle $\theta$ is obtained by summing all the partial-wave contributions
\begin{align}
f_{C}(k,\theta) 
&= \sum_{l}
(2l+1)f_{C,l}(k) P_{l}(\cos\theta) \\ 
&= -\frac{[\eta(k)]^{2}}{2k_{C}}\frac{e^{2i\sigma_{0}(k)}}{\sin^{2}\left(\frac{\theta}{2}\right)}\exp\left[- i\eta(k) \log\left(\sin^{2}\frac{\theta}{2}\right) \right],
\label{eq:fC}
\end{align}
where $\sigma_{0}(k) = \arg\Gamma[1 + i\eta(k)]$ is the $s$-wave Coulomb phase shift. This scattering amplitude leads to the well-known cross section for Rutherford scattering.

In the present study, we focus on the $s$-wave scattering:
\begin{align}
f_{C}(k) &= f_{C,0}(k)=\frac{e^{2i\sigma_{0}(k)} - 1}{2ik} .
\label{eq:fC-swave}
\end{align}
Hereafter, we suppress the $k$ dependence of $\sigma_{0}$ and $\eta$ for brevity. To search for poles in the complex momentum plane in the later discussion, we rewrite $f_{C}(k)$ in a suitable form for analytic continuation. Note that $\sigma_{0} =\arg\Gamma(1 + i\eta)$ is not an analytic function because it involves taking the complex phase (argument) of the gamma function. We thus first rewrite $\sigma_{0}$ in an analytic form. From the integral form of the gamma function, it is clear that
\begin{align}
\Gamma(1 + i\eta) &= [\Gamma(1 - i\eta)]^{*},
\end{align}
for real $\eta$. Using this relation and the polar representation of the gamma function, we obtain 
\begin{align}
e^{2i\sigma_{0}} &= \frac{\Gamma(1 + i\eta)}{\Gamma(1 - i\eta)}.
\label{eq:gamma-exp}
\end{align}
Since the gamma function is analytic, this expression holds for arbitrary complex values of $\eta$. By substituting this expression into Eq.~\eqref{eq:fC-swave}, the pure-Coulomb scattering amplitude is rewritten in terms of gamma functions:
\begin{align}
f_{C}(k) &= \frac{\eta}{2ik_{C}}\left[\frac{\Gamma(1 + i\eta)}{\Gamma(1 - i\eta)} - 1\right].
\label{eq:fC-swave-analytic}
\end{align}
This form will be used to discuss the pole positions of the pure-Coulomb scattering amplitude in Sec.~\ref{subsec:pole-condition}. 

\subsubsection{Coulomb-distorted short-range scattering amplitude}

The second term $f_{CS}(k)$ in Eq.~\eqref{eq:decomposition} corresponds to the Coulomb-distorted short-range amplitude. In order to write down $f_{CS}(k)$, we define two functions of $\eta$ which encode the Coulomb contributions in the amplitude. The first is the Sommerfeld factor $C_{\eta}^{2}$~\cite{Konig:2012prq}:
\begin{align}
C_{\eta}^{2} &= \frac{2\pi\eta}{e^{2\pi \eta} - 1},
\label{eq:C-eta-2}
\end{align}
and the second is the $H$ function~\cite{Kok:1982dr,Kong:1999sf}:
\begin{align}
H(\eta) &= \psi(i\eta) + \frac{1}{2i\eta} - \log[i\eta\ \operatorname{sgn}(k_{C})],
\label{eq:H} \\
&= \psi(1 + i\eta) - \frac{1}{2i\eta} - \log[i\eta\ \operatorname{sgn}(k_{C})].
\label{eq:H-ik}
\end{align}
Here, $\psi(z)$ is the digamma function:
\begin{align}
\psi(z) &= \frac{d}{dz}\log[\Gamma(z)].
\end{align}
The expression in Eq.~\eqref{eq:H-ik} is obtained by using the property of the digamma function
\begin{align}
\psi(z + 1) &= \psi(z) + \frac{1}{z}.
\end{align}

Because the $s$-wave short-range interaction is introduced in Sec.~\ref{subsec:Hamiltonian}, we focus on the $s$-wave component of $f_{CS}(k,\theta)$. The $s$-wave Coulomb-distorted short-range amplitude $f_{CS}(k)$ has no angular dependence, and is computed by solving the LS equation with the Coulomb-modified Green function $G_{C}$ instead of $G_{0}$~\cite{Kong:1998sx,Kong:1999sf,Ando:2007fh,Higa:2008dn}. To remove the divergences arising from the momentum integrals in the calculation, we introduce dimensional regularization. Then the amplitude $f_{CS}(k)$ is obtained as~\cite{Higa:2008dn}
\begin{align}
f_{CS}(k) &= C_{\eta}^{2}e^{2i\sigma_{0}}
\left[\sigma\frac{2\pi\Delta^{(R)}}{\mu g_{0}^{2}} - \sigma\frac{\pi}{\mu^{2}g_{0}^{2}}k^{2} - 2k_{C}H(\eta)\right]^{-1}.
\label{eq:TCS-bare}
\end{align}
Here, the constant $\Delta^{(R)}$ is the Coulomb-renormalized mass parameter, which is given by
\begin{align}
\Delta^{(R)} &= \nu_{0}(\Lambda) - \sigma\frac{\mu g_{0}^{2}}{2\pi} \left\{
\frac{\Lambda}{D - 3} \right.\nonumber \\
\quad &\quad + \left. 2k_{C}\left[\frac{1}{D - 4} - \log\left(\frac{\sqrt{\pi}\Lambda}{2k_{C}}\right) - 1 + \frac{3}{2}C_{E}\right]\right\},
\label{eq:Delta-R}
\end{align}
with the renormalization scale $\Lambda$, the space-time dimension $D$, and the Euler constant $C_{E} = -\psi(1)$. 

Here we derive the relation between the model parameters and the scattering observables by comparing Eq.~\eqref{eq:TCS-bare} with the effective range expansion with the Coulomb contribution~\cite{Higa:2008dn}. In general, $f_{CS}(k)$ can be written using the Coulomb-corrected phase shift $\delta_{0}^{C}$~\cite{Kong:1999sf,Higa:2008dn}:
\begin{align}
f_{CS}(k) &= \frac{C_{\eta}^{2}e^{2i\sigma_{0}}}{k(\cot \delta_{0}^{C} - i)}.
\label{eq:fCS-general}
\end{align}
At low energies with small $|k|$, the denominator of $f_{CS}(k)$ can be expanded as~\cite{Landau:1944jqr,Bethe:1949yr,Higa:2008dn}
\begin{align}
f_{CS}(k) &= C_{\eta}^{2}e^{2i\sigma_{0}}\left[-\frac{1}{a^{C}_{s}} + \frac{r^{C}_{e}}{2}k^{2} + \mathcal{O}(k^{4}) - 2k_{C}H(\eta) \right]^{-1}.
\label{eq:ERE-C}
\end{align}
Here, $a^{C}_{s}$ and $r^{C}_{e}$ stand for the Coulomb scattering length and the Coulomb effective range, respectively, which characterize the low-energy behavior of $f_{CS}(k)$. By comparing Eq.~\eqref{eq:ERE-C} with Eq.~\eqref{eq:TCS-bare}, we find that $a^{C}_{s}$ and $r^{C}_{e}$ are given by
\begin{align}
a^{C}_{s} &= -\sigma\frac{\mu g_{0}^{2}}{2\pi \Delta^{(R)}},
\label{eq:as-C}\\
r^{C}_{e} &= -\sigma\frac{2\pi}{\mu^{2}g_{0}^{2}}.
\label{eq:re-C}
\end{align}
From this expression, we find that the effective range $r_{e}^{C}$ always becomes negative (positive) for $\sigma = 1$ ($\sigma = -1$). On the other hand, by varying $\Delta^{(R)}$, the scattering length $a_{s}^{C}$ can take arbitrary values ($-\infty < a_{s}^{C}< \infty$), irrespective of $\sigma$. In terms of these scattering observables $a^{C}_{s}$ and $r^{C}_{e}$, the scattering amplitude in the EFT model $f_{CS}(k)$ in Eq.~\eqref{eq:TCS-bare} is written in terms of the observables
\begin{align}
f_{CS}(k) &= C_{\eta}^{2}e^{2i\sigma_{0}}\left[-\frac{1}{a^{C}_{s}} + \frac{r^{C}_{e}}{2}k^{2} - 2k_{C}H(\eta) \right]^{-1}.
\label{eq:fCS-renorm}
\end{align}
In this way, we find that $f_{CS}(k)$ in this model is written in terms of $a_{s}^{C}$ and $r_{e}^{C}$ without higher-order corrections at $\mathcal{O}(k^{4})$. This feature is similar to the resonance model for the short-range interaction in Ref.~\cite{Braaten:2007nq}.

\subsubsection{Scattering amplitude in the absence of Coulomb or short-range interactions}
\label{subsubsec:behavior-of-amplitude}

To understand how the Coulomb and short-range interactions individually contribute to the scattering amplitude, we examine the behavior of the full amplitude $f(k)$ in the two limiting cases where each interaction is switched off in turn.

First, we focus on the $g_{0} \to 0$ limit where the short-range interaction vanishes. In this case, the bare field $\phi$ decouples from the scattering states, leaving only the Coulomb interaction. In Eq.~\eqref{eq:TCS-bare}, the Coulomb-distorted short-range amplitude $f_{CS}(k)$ vanishes, and the system is therefore described solely by the pure Coulomb amplitude $f_{C}(k)$ in Eq.~\eqref{eq:fC-swave}. As we show later, in this case, an infinite number of Coulomb eigenstates appear near the threshold. 

We then consider the opposite limit in which the Coulomb interaction vanishes, defined by $\alpha\to 0$. From Eq.~\eqref{eq:aB}, this corresponds to the $k_{C} \to 0$ limit, or equivalently, the $a_{B} \to \infty$ limit. Then, for fixed momentum $k$, $\eta\to 0$ in Eq.~\eqref{eq:eta}, and hence $C_{\eta}^{2} \to 1$ and $\sigma_{0} \to \arg\Gamma(1) = 0$ as indicated in Eqs.~\eqref{eq:C-eta-2} and~\eqref{eq:sigma-l}. 
In this limit, the pure Coulomb amplitude $f_{C}(k)$ in Eq.~\eqref{eq:fC-swave} vanishes, while $f_{CS}(k)$ remains finite. Furthermore, Eq.~\eqref{eq:H-ik} indicates that 
\begin{align}
2k_{C}H(\eta) \to ik \quad (k_{C} \to 0).
\end{align}
This $ik$ term is required by the unitarity of the $s$-matrix~\cite{Taylor}, regardless of the specific form of the amplitude. Consequently, the Coulomb-distorted amplitude $f_{CS}(k)$ reduces to
\begin{align}
f_{CS}(k) \to \left[\sigma\frac{2\pi\Delta^{(R)}_{S}}{\mu g_{0}^{2}} - \sigma\frac{\pi}{\mu^{2}g_{0}^{2}}k^{2} - ik\right]^{-1} \quad (\alpha \to 0),
\label{eq:f-CS-SR}
\end{align}
where $\Delta^{(R)}_{S} = \nu_{0}(\Lambda) - \sigma\mu g_{0}^{2}\Lambda/[2\pi(D - 3)]$. This coincides with the scattering amplitude obtained from the short-range Hamiltonian~\eqref{eq:H-free-SR} and \eqref{eq:H-int-SR}~\cite{Kinugawa:2022fzn}. From Eq.~\eqref{eq:f-CS-SR}, we find that the scattering length $a_{s}$ and the effective range $r_{e}$ in this limit are given by
\begin{align}
a_{s} &= -\sigma\frac{\mu g_{0}^{2}}{2\pi \Delta^{(R)}_{S}},
\label{eq:as-SR}\\
r_{e} &= -\sigma\frac{2\pi}{\mu^{2}g_{0}^{2}}.
\label{eq:re-SR}
\end{align}
Equation~\eqref{eq:as-SR} is obtained by replacing $\Delta^{(R)}$ with $\Delta_{S}^{(R)}$ in Eq.~\eqref{eq:as-C}. Meanwhile, the expression for the effective range remains unchanged regardless of the presence of the Coulomb interaction~\cite{Kong:1999sf}. From Eqs.~\eqref{eq:as-SR} and \eqref{eq:re-SR}, we find that Eq.~\eqref{eq:fCS-renorm} reduces to the ERE in systems with the short-range interaction:
\begin{align}
f_{CS}(k) \to \left[-\frac{1}{a_{s}} + \frac{r_{e}}{2}k^{2} - ik\right]^{-1} \quad (\alpha \to 0).
\label{eq:bsc-sr}
\end{align}
In the present model, by comparing Eqs.~\eqref{eq:as-C} and \eqref{eq:re-C} with Eqs.~\eqref{eq:as-SR} and \eqref{eq:re-SR}, the scattering parameters with the Coulomb interaction ($a_{s}^{C}$ and $r_{e}^{C}$) can be related to those without the Coulomb interaction ($a_{s}$ and $r_{e}$). However, in general, such relations cannot be obtained without specifying a particular model.

The above discussion indicates that the short-range interaction becomes weaker as $g_{0}$ decreases. From Eq.~\eqref{eq:re-C}, the system with the weak short-range interaction corresponds to that with the large magnitude of the effective range $|r_{e}^{C}|$. Furthermore, the Coulomb interaction becomes weaker as $a_{B}$ increases, which is natural since $a_{B}$ in Eq.~\eqref{eq:aB} is proportional to $1/\alpha$. Thus, the ratio $|r_{e}^{C}|/a_{B}$ serves as a measure of the strength of the Coulomb interaction relative to the short-range interaction. We summarize its correspondence to the strength of the Coulomb interaction in Table~\ref{tab:limits-sum}.

\begin{table}
 \caption{Relation between the quantities and the strength of the interactions.\label{tab:limits-sum}}
  \begin{ruledtabular}
  \begin{tabular}{cc}
    Parameters & Strength of interaction \\ \hline 
    $|r_{e}^{C}|/a_{B} \gg 1$ & strong Coulomb interaction \\
    $|r_{e}^{C}|/a_{B} \ll 1$ & weak Coulomb interaction \\
  \end{tabular}
  \end{ruledtabular}
\end{table}

\subsection{Pole condition and self-energy}
\label{subsec:pole-condition}

In the above discussion, we have introduced the pure Coulomb amplitude $f_{C}(k)$ and the Coulomb-distorted short-range amplitude $f_{CS}(k)$. These amplitudes encode information on the bound state and the resonance through their analytic structure in the complex momentum plane. In the following, we determine the eigenmomenta by analyzing the pole condition of the amplitude.

\subsubsection{Pole condition}
\label{subsubsec:pole-condition}

As shown in Eq.~\eqref{eq:decomposition}, the full scattering amplitude $f(k)$ can be decomposed into two parts, the pure Coulomb amplitude $f_{C}(k)$ and the Coulomb-distorted short-range amplitude $f_{CS}(k)$. First, we examine the pure-Coulomb scattering amplitude $f_{C}(k)$. In Eq.~\eqref{eq:fC-swave-analytic}, the poles arise solely from the factor $\Gamma(1 + i\eta)$. Because the gamma function $\Gamma(z)$ has poles at $z = 0, -1, -2, \ldots$, the poles of $\Gamma(1 + i\eta) = \Gamma(1 + ik_{C}/k)$ are located at 
\begin{align}
k = k_{n} \equiv -i\frac{k_{C}}{n} \quad (n = 1,2,3, \ldots).
\label{eq:Coulomb-poles}
\end{align}
For an attractive Coulomb interaction ($k_{C} < 0$), Eq.~\eqref{eq:fC-swave-analytic} thus possesses an infinite number of poles located at $k = i|k_{C}|/n$ on the positive imaginary $k$ axis, which are associated with bound states. In contrast, for a repulsive Coulomb interaction with $k_{C} > 0$, $f_{C}(k)$ also has an infinite number of poles, located at $k = -i|k_{C}|/n$ on the negative imaginary axis, which correspond to virtual states~\cite{Mochizuki:2024dbf}. These poles correspond to the well-known bound and virtual states generated solely by the Coulomb interaction without the short-range interaction.

We then focus on the poles of the Coulomb-distorted short-range amplitude $f_{CS}(k)$. It can be shown that $f_{CS}(k)$ has poles at the positions given in Eq.~\eqref{eq:Coulomb-poles}, and that their residues are equal in magnitude but opposite in sign to those of $f_{C}(k)$. As a result, these pole contributions cancel in the full amplitude in Eq.~\eqref{eq:decomposition}. Consequently, the pole condition for $f_{CS}(k)$ in Eq.~\eqref{eq:TCS-bare} is given by 
\begin{align}
\sigma\frac{2\pi\Delta^{(R)}}{\mu g_{0}^{2}} - \sigma\frac{\pi}{\mu^{2}g_{0}^{2}}k^{2} - 2k_{C}H(\eta) &= 0.
\label{eq:fCS-inv}
\end{align}

From $f_{CS}(k)$ written in terms of the observables~\eqref{eq:fCS-renorm}, the pole condition~\eqref{eq:fCS-inv} is also expressed by the Coulomb scattering length $a_{s}^{C}$~\eqref{eq:as-C} and the Coulomb effective range $r_{e}^{C}$~\eqref{eq:re-C}~\cite{Domcke:1983zz}:
\begin{widetext}
\begin{align}
-\frac{a_{B}}{a_{s}^{C}} + \frac{r_{e}^{C}}{2a_{B}}(a_{B}k)^{2} - ia_{B}k - 2\, \operatorname{sgn}(Z_{1}Z_{2})\left \{\log(-ia_Bk) + \psi\left(1 + \frac{i\ \operatorname{sgn}(Z_{1}Z_{2})}{a_Bk}\right)\right\} &= 0
\label{eq:pole-condition-ERE}
\end{align}
\end{widetext}
In later sections, we will numerically solve the pole condition~\eqref{eq:pole-condition-ERE} for both attractive and repulsive Coulomb cases. 


\subsubsection{Classification of poles}
\label{subsec:poles}

To classify the poles, we first discuss the analytic structure of the scattering amplitude $f_{CS}(k)$. The logarithmic term in Eq.~\eqref{eq:pole-condition-ERE} gives rise to the multi-valued structure of the scattering amplitude $f_{CS}(k)$. For a complex variable $z$, the logarithm is defined as 
\begin{align}
\log(z) = \log|z| + i(\arg(z) + 2n\pi), \quad n = 0, \pm 1, \pm 2, \ldots
\label{eq:log-def}
\end{align}
where the argument of $z$ is taken in the range $-\pi \leq \arg(z) < \pi$, and the branch cut runs from the origin to $-\infty$. Thus, $\log(-ia_{B}k)$ in Eq.~\eqref{eq:pole-condition-ERE} has a branch cut along the negative imaginary axis in the complex momentum plane. This feature is analogous to the Yukawa potential, whose scattering amplitude also has a branch cut along the imaginary $k$ axis for $\Im\, k \leq -\mu/2$ with $\mu$ being the mass of the exchanged particle~\cite{Taylor}. A similar logarithmic cut in the momentum plane appears in the three-body scattering amplitude, as studied in the context of Efimov physics~\cite{Hyodo:2013zxa,Dawid:2023kxu}.

The branch cut of the logarithmic term generates an infinite number of Riemann sheets in the complex momentum plane. The Riemann sheet with $n=0$ is the sheet adjacent to the physical scattering on the positive real $k$ axis. Therefore, we mainly discuss the poles in the Riemann $n = 0$ sheet throughout this study. Bound-state poles are located on the positive imaginary axis of the $k$ plane and are the only poles that appear in the upper half plane. In the lower half plane, a virtual-state pole cannot lie on the negative imaginary axis because of the branch cut, in contrast to the case without the Coulomb interaction. We classify the poles in the lower half plane according to the argument of the momentum, $\theta_{k}=\arg(k)$. Poles in the fourth quadrant are closer to the physical scattering region on the positive real axis and are therefore regarded as being relevant for physical observables. Among them, poles with $-\pi/4 \leq \theta_{k} < 0$ represent states with a positive real part of the energy and are thus interpreted as resonances. Poles in the region $-\pi/2 < \theta_{k} < -\pi/4$ correspond to solutions with negative energy and a finite imaginary part, which we refer to here as virtual states. In the third quadrant of the momentum plane, poles that are conjugate to the resonances and virtual states appear. These are referred to as anti-resonances ($-\pi < \theta_{k} \leq -3\pi/4$) and anti-virtual states ($-3\pi/4 < \theta_{k} < -\pi/2$), respectively. The classification of the poles in the complex $k$ plane is summarized in Table~\ref{tab:poles}.
 
\begin{table}
 \caption{Classification of poles of the amplitude $f_{CS}(k)$ in the $n=0$ Riemann sheet of the complex momentum plane, according to the argument $\theta_k=\arg(k)$. \label{tab:poles}} 
\begin{ruledtabular}
  \begin{tabular}{ll} 
   State & Argument $\theta_{k}$ \\
   \hline 
   Bound state & $\theta_{k} = \pi/2$ (positive imaginary axis) \\
   Resonance & $-\pi/4 \leq \theta_{k} < 0$ \\
   Virtual state& $-\pi/2<\theta_{k} < -\pi/4$ \\
   Anti-virtual state & $-3\pi/4 < \theta_{k} < -\pi/2 $ \\
   Anti-resonance & $-\pi < \theta_{k} \leq -3\pi/4$ \\
 \end{tabular}
\end{ruledtabular}
\end{table} 

In preparation for the numerical investigation of the eigenmomenta in Secs.~\ref{sec:negative-re} and \ref{sec:positive-re}, we consider the behavior of the poles determined by Eq.~\eqref{eq:pole-condition-ERE} in the limit $|a_{B}/a_{s}^{C}| \to \infty$ with fixed $a_{B}$ and $r_{e}^{C} < 0$. From Eq.~\eqref{eq:as-C} and \eqref{eq:re-C}, this limit corresponds to the case with $|\Delta^{(R)}| \to \infty$ with fixed $g_{0}^{2}$. In this case, we may regard the bare state as being almost decoupled from the Coulomb eigenstates. Thus, the poles can be classified into two categories: those associated with the pure short-range interaction and those with the pure Coulomb interaction. The former poles appear in the region where $|k|$ is sufficiently large to neglect the fourth term in Eq.~\eqref{eq:pole-condition-ERE}. In this region, the eigenmomenta are determined only by $a_{s}^{C}$ and $r_{e}^{C}$ as 
\begin{align}
k^{\pm}& \to \frac{i}{r_{e}^{C}}\pm\frac{1}{r_{e}^{C}}\sqrt{\frac{2r_{e}^{C}}{a_{s}^{C}}-1+i0^{+}}
\label{eq:k-pm} \\
&
\to 
\begin{cases}
\dfrac{i}{r_{e}^{C}}\mp \infty & a_{B}/a^{C}_{s}\to +\infty \\
\mp i\infty & a_{B}/a^{C}_{s}\to -\infty
\end{cases} .
\end{align}
Equation~\eqref{eq:k-pm} gives the pole positions for a pure short-range interaction~\cite{Hyodo:2013iga}. For $a_{B}/a_{s}^{C}\to -\infty$, these solutions describe a deeply bound state and a virtual state. In the $a_{B}/a_{s}^{C}\to \infty$ limit, the imaginary part of $k$ remains fixed at $1/r_{e}^{C}$ and the real part becomes large, leading to a resonance and an anti-resonance.
On the other hand, in the small-$|k|$ region in the $|a_{B}/a_{s}^{C}| \to \infty$ limit, the solutions of Eq.~\eqref{eq:pole-condition-ERE} correspond to the poles of the digamma function. These poles correspond to the Coulomb bound or virtual states discussed around Eq.~\eqref{eq:Coulomb-poles}. Thus, in the $|a_{B}/a_{s}^{C}| \to \infty$ limit, the poles in the system are completely separated into those generated by the pure Coulomb interaction in the small-$|k|$ region and those generated by the short-range interaction in the large-$|k|$ region.

\subsubsection{Self-energy}
\label{subsubsec:self-energy}

Finally, in preparation for the calculation of the compositeness in the next section, we present the expression for the self-energy $\Sigma(E)$ obtained from the pole condition. The pole condition~\eqref{eq:fCS-inv} can be rewritten in terms of the eigenenergy $E = k^{2}/(2\mu)$ as
\begin{align}
E + \sigma\frac{\mu g_{0}^{2}k_{C}}{\pi}H(\eta) - \Delta^{(R)} = 0.
\label{eq:pole-condition}
\end{align}
In terms of the self-energy $\Sigma(E)$, the same pole condition can be written as
\begin{align}
E - \nu_{0}(\Lambda) - \Sigma(E) = 0.
\label{eq:pole-condition-Sigma}
\end{align}
That is, the self-energy describes how the bare field $\phi$ with bare energy $\nu_{0}$ is dressed through its coupling to the scattering states. By comparing Eq.~\eqref{eq:pole-condition} and Eq.~\eqref{eq:pole-condition-Sigma}, the self-energy in this model is given by
\begin{align}
\Sigma(E) &= -g_{0}^{2} \sigma\frac{\mu k_{C}}{\pi}H(\eta)+C(\Lambda),
\label{eq:Sigma}
\end{align}
where $C(\Lambda)$ is an energy-independent constant. Under renormalization, the divergent part of $C(\Lambda)$ is absorbed into $\nu_{0}(\Lambda)$, while a finite constant contribution, which depends on the renormalization scale $\Lambda$, remains in the self-energy. As we will show later, the compositeness is determined solely by the energy derivative of the self-energy. Therefore, in the present regularization, the compositeness is independent of the renormalization scale.

\subsection{Compositeness}
\label{subsec:compositeness}

Here, we introduce the compositeness as a quantitative measure to investigate the internal structure of the eigenstates. To this end, we define the free scattering states $\ket{\bm{p}}$ and a discrete bare state $\ket{\phi}$, which are eigenstates of the free Hamiltonian~\eqref{eq:H-free-SR}:
\begin{align}
H_{0} \ket{\bm{p}} &= \frac{\bm{p}^{2}}{2\mu}\ket{\bm{p}},\\
H_{0} \ket{\phi} &= \nu_{0}\ket{\phi}.
\end{align}
These eigenstates are normalized as
\begin{align}
\braket{\bm{p}'|\bm{p}} &= (2\pi)^{3} \delta(\bm{p}' - \bm{p}), \quad \braket{\phi|\phi} = \sigma.
\end{align}
That is, $\ket{\phi}$ is a negative-norm state if $\sigma = -1$. The completeness relation is 
\begin{align}
1 = \int \frac{d^{3}p}{(2\pi)^{3}} \ket{\bm{p}}\bra{\bm{p}} + \sigma \ket{\phi}\bra{\phi},
\label{eq:completeness}
\end{align}
where the bare-state term also contains the factor $\sigma$.

\subsubsection{Compositeness of bound states}
If the system has a bound state $\ket{B}$ with the binding energy $B$, it satisfies the Schr\"odinger equation:
\begin{align}
H \ket{B} = (H_{0} + H_{S} + H_{C}) \ket{B} = -B \ket{B}.
\label{eq:bound-state}
\end{align}
We express $\ket{B}$ as a superposition of $\ket{\bm{p}}$ and $\ket{\phi}$ using the completeness relation in Eq.~\eqref{eq:completeness}. Then, the compositeness $X$ (the elementarity $Z$) is defined through the overlap of $\ket{\bm{p}}$ ($\ket{\phi}$) and the bound state $\ket{B}$~\cite{Hyodo:2013nka,vanKolck:2022lqz,Kinugawa:2024crb}:
\begin{align}
X &= \int \frac{d^{3}p}{(2\pi)^{3}}|\braket{\bm{p}|B}|^{2},
\label{eq:X-def}\\
Z &= \sigma|\braket{\phi|B}|^{2}, \label{eq:Z-def}\\
X + Z &= 1.
\label{eq:normalize}
\end{align}
When $\sigma = +1$, $X$ and $Z$ are positive by definition. Furthermore, Eq.~\eqref{eq:normalize} ensures that $X$ and $Z$ are normalized and bounded by unity. That is, the compositeness and elementarity take values in the range 
\begin{align}
0 \leq X \leq 1, \quad 0 \leq Z \leq 1 \quad (\sigma = +1).
\end{align}
Based on these properties, the compositeness $X$ can be interpreted as the probability of finding the molecular component in the wavefunction, while the elementarity $Z$ corresponds to that of the compact bare state. On the other hand, when $\sigma = -1$, the elementarity $Z$ is always negative according to Eq.~\eqref{eq:Z-def}, and Eq.~\eqref{eq:normalize} indicates that the compositeness exceeds unity:
\begin{align}
X \geq 1, \quad Z \leq 0 \quad (\sigma = -1).
\label{eq:negative-Z}
\end{align}
In this case, $X$ and $Z$ cannot be regarded as probabilities. The fact that the compositeness exceeds unity can be understood as a consequence of the negative-norm bare state, $\braket{\phi|\phi} = -1$, used in the definition of $Z$. We will discuss how to interpret the compositeness for $\sigma = -1$ in a later section.

\subsubsection{Expressions for compositeness}
\label{subsec:expression-X}

For practical calculations, it is convenient to evaluate the compositeness using various equivalent expressions~\cite{Hyodo:2013nka,Kinugawa:2024crb}. Among the existing formulas, the expression in terms of the effective interaction and the loop function~\cite{Sekihara:2014kya,Kamiya:2015aea,Kamiya:2016oao} has frequently been used in previous studies. However, this expression is not directly applicable to the present system, as the Coulomb interaction is not separable. Instead, we employ an alternative expression given in terms of the self-energy $\Sigma(E)$~\cite{Hyodo:2014bda,Kinugawa:2024crb}, which can be applied to systems with non-separable Coulomb plus short-range interactions:
\begin{align}
X &= -\left. \frac{\frac{d}{dE}\Sigma(E)}{1 - \frac{d}{dE}\Sigma(E)} \right|_{E = -B}, 
\label{eq:X} \\
Z &= \left. \frac{1}{1 - \frac{d}{dE}\Sigma(E)} \right|_{E = -B}.
\label{eq:Z}
\end{align}
In the present model, the energy derivative of $\Sigma(E)$ in Eq.~\eqref{eq:Sigma} is calculated as
\begin{align}
\frac{d}{dE}\Sigma(E) &= -\sigma\frac{g_{0}^{2}\mu k_{C}}{\pi}\frac{d}{dE}H(\eta) \nonumber \\
&= \sigma\frac{g_{0}^{2}\mu^{2}}{2\pi}\frac{2k_{C}^{2}}{k^{3}}\left[i\psi_{1}(i\eta) - \frac{1}{2i\eta^{2}} - \frac{1}{\eta} \right] \nonumber \\
&= -\sigma\frac{1}{r_{e}^{C}k}\left[2i\eta^{2}\psi_{1}(i\eta) + i - 2\eta \right],
\label{eq:Sigma-prime}
\end{align}
with the trigamma function $\psi_{1}(z) = \frac{d^{2}}{dz^{2}}\log[\Gamma(z)]$. 

Equation~\eqref{eq:Sigma-prime} indicates that the compositeness $X$ can be written solely in terms of the observables
\begin{align}
X &= \left[1 - \frac{r_{e}^{C}}{R^{C}}\right]^{-1}, 
\label{eq:wbr}
\end{align}
where $R^{C}$ is defined as
\begin{align}
R^{C} 
 &= \left.r_{e}^{C}\frac{d}{dE}\Sigma(E)\right|_{E = -B} 
 \label{eq:sigma-p}\\
 &= -\sigma\frac{1}{k_{h}}\left[2i\left(\frac{k_{C}}{k_{h}}\right)^{2}\psi_{1}\left(i\frac{k_{C}}{k_{h}}\right)+ i - 2\left(\frac{k_{C}}{k_{h}}\right) \right]
 \label{eq:R-C-2}\\
 &= \sigma\frac{i}{k_{h}} - \sigma\frac{i}{k_{h}}\left[2\left(\frac{k_{C}}{k_{h}}\right)^{2}\psi_{1} \left(i\frac{k_{C}}{k_{h}} + 1\right) + 2i\left(\frac{k_{C}}{k_{h}}\right)\right].
\label{eq:R-C}
\end{align}
Here, $k_{h}$ is the eigenmomentum $k_{h} = i\kappa_{h}$ with $\kappa_{h} = \sqrt{2\mu B}$.

In a purely short-range system, the compositeness is related to the effective range $r_e$ and the bound-state radius $R = 1/\kappa_h$ through the weak-binding relation~\cite{Weinberg:1965zz,Kamiya:2015aea,Kamiya:2016oao,Kinugawa:2022fzn}
\begin{align}
X &= \left[1 - \frac{r_{e}}{R}\right]^{-1}.
\label{eq:wbr-SR}
\end{align}
By comparing Eqs.~\eqref{eq:wbr} and \eqref{eq:wbr-SR}, $R^{C}$ can be interpreted as the radius of the bound-state wavefunction in the presence of the Coulomb interaction. In fact, when $\sigma = +1$, the first term in Eq.~\eqref{eq:R-C} corresponds to the radius of the bound state $R = i/k_{h}$ in the purely short-range system. This suggests that the second term in Eq.~\eqref{eq:R-C} gives the Coulomb contribution to the radius. 

Using the pole condition for the eigenmomentum~\eqref{eq:pole-condition-ERE}, the compositeness $X$ in Eq.~\eqref{eq:wbr} can be further rewritten in terms of the scattering length $a_{s}^{C}$, the effective range $r_{e}^{C}$, and the Bohr radius $a_{B}$. That is, the compositeness of near-threshold states is independent of model details. This shows a kind of universality governed by $a_{s}^{C}$, $r_{e}^{C}$, and $a_{B}$, although the usual low-energy universality does not hold in the presence of the Coulomb interaction. In this sense, Eq.~\eqref{eq:wbr} serves as the weak-binding relation for systems with Coulomb plus short-range interactions.

Equation~\eqref{eq:wbr} shows that $X \to 1$ when $R^{C} \to \infty$ with finite $r_{e}^{C}$. The divergence of $R^{C}$ occurs at the poles of the trigamma function\footnote{$R^{C}$ is finite in the $k_{h} \to 0$ limit as we will see below.}, which are located at the eigenmomenta of the pure Coulomb bound states [$k_{n}$ in Eq.~\eqref{eq:Coulomb-poles}]. When a bound state exists near the pure Coulomb level ($k_{h} \sim k_{n}$), the compositeness is expected to be $X \sim 1$. Such a pole is realized when $|a_{B}/a_{s}^{C}| \to \infty$ as discussed in Sec.~\ref{subsec:poles}. This is consistent with the expectation that the pure Coulomb bound states are completely composite. We will confirm this numerically in Sec.~\ref{subsubsec:X-att-neg}.

Next, we examine the compositeness of the bound state exactly at the threshold. The quantity $k_{C}R^{C}$ can be rewritten in terms of dimensionless quantity $\tilde{\kappa} = \kappa_{h}/k_{C}$ as
\begin{align}
k_{C}R^{C} &= \frac{2}{\tilde{\kappa}^{3}}\psi_{1}\!\left(\frac{1}{\tilde{\kappa}}\right) - \frac{1}{\tilde{\kappa}} - \frac{2}{\tilde{\kappa}^{2}}.
\label{eq:R-kappa}
\end{align}
To evaluate $\tilde{\kappa}\to 0$ or equivalently $1/\tilde{\kappa}\to\infty$, we use the asymptotic expansion of the trigamma function for $|z|\to\infty$,
\begin{align}
\psi_{1}(z) &= \frac{1}{z} + \frac{1}{2z^{2}} + \frac{1}{6z^{3}} - \frac{1}{30z^{5}} + \mathcal{O}\!\left(\frac{1}{z^{7}}\right).
\end{align}
From Eq.~\eqref{eq:R-kappa}, we then obtain
\begin{align}
k_{C}R^{C} \to \frac{1}{3} \quad (\kappa_{h}\to 0).
\end{align}
The compositeness of the bound state at the threshold then becomes
\begin{align}
X \to \frac{1}{1 - 3k_{C}r_{e}^{C}}
 = \frac{1}{1 - 3\, \operatorname{sgn}(Z_{1}Z_{2})r_{e}^{C}/a_{B}}
\quad (B\to 0).
\label{eq:X-B=0}
\end{align}
Thus, the value of $X$ does not approach unity in the $B \to 0$ limit. In fact, $X$ at the threshold is determined by the interplay between the Bohr radius $a_{B}$ and the effective range $r_{e}^{C}$. When $a_{B}$ is large, the contribution of the Coulomb interaction is relatively smaller than that of the short-range interaction as shown in Table~\ref{tab:limits-sum}, and the compositeness $X$ takes a value close to unity. In the $a_{B}\to\infty$ limit where the Coulomb interaction is switched off, $X$ becomes exactly unity, which is in line with the short-range universality~\cite{Hyodo:2013iga}.

In contrast, as the pole moves away from the threshold, the influence of the scattering states becomes weaker. In fact, Eq.~\eqref{eq:R-C-2} shows that $R^{C}$ decreases for sufficiently large $|k_{h}|$. Consequently, the compositeness $X$ becomes smaller, because the overlap $\braket{\bm{p}|B}$ in Eq.~\eqref{eq:X-def} decreases. In the $|k_{h}|\to\infty$ limit with fixed $r_{e}^{C}$, Eq.~\eqref{eq:R-C-2} gives $R^{C}\to 0$, and Eq.~\eqref{eq:wbr} leads to $X\to 0$.

\subsubsection{Compositeness of resonances}
So far, we have considered the compositeness of stable bound states. In the following, we will also consider resonances located above the threshold, such as ${}^{8}$Be and the $\Omega_{ccc}^{++}\Omega_{ccc}^{++}$ dibaryon mentioned in the introduction. The extension of the compositeness to unstable resonances is achieved by introducing the Gamow vector~\cite{Berggren:1968zz,Bohm:1981pv,Kukulin,Bohm:2001,delaMadrid:2002cz,Moiseyev,delaMadrid:2024izo}. Using the Gamow bra vector $\bra{\tilde{R}}$, the compositeness of a resonance $\ket{R}$ is defined as~\cite{Hyodo:2013nka,Aceti:2014ala,Kinugawa:2024crb}
\begin{align}
X &= \int \frac{d^{3}p}{(2\pi)^{3}} \braket{\tilde{R}|\bm{p}}\braket{\bm{p}|R} = \int \frac{d^{3}p}{(2\pi)^{3}} [\braket{\bm{p}|R}]^{2},
\label{eq:X-def-Gamow} \\
Z &= \sigma \braket{\tilde{R}|\phi}\braket{\phi|R} = \sigma [\braket{\phi|R}]^{2}.
\label{eq:Z-def-Gamow}
\end{align}
Here, $X$ and $Z$ are obtained from the square of the wavefunctions $\braket{\bm{p}|R}$ and $\braket{\phi|R}$, not from their absolute value squared. Since resonances are normalized as $\braket{\tilde{R}|R} = 1$, the compositeness and the elementarity satisfy $X + Z = 1$ even for resonances.

Because the wavefunctions $\braket{\bm{p}|R}$ and $\braket{\phi|R}$ are in general complex, $X$ in Eq.~\eqref{eq:X-def-Gamow} and $Z$ in Eq.~\eqref{eq:Z-def-Gamow} also become complex. This complicates the probabilistic interpretation of the compositeness of resonances. To extract information about the internal structure from the complex-valued compositeness, several prescriptions have been proposed~\cite{Aceti:2012dd,Xiao:2012vv,Aceti:2014ala,Hyodo:2013iga,Hyodo:2013nka,Sekihara:2013sma,Kamiya:2015aea,Sekihara:2015gvw,Kamiya:2016oao,Matuschek:2020gqe,Kinugawa:2024kwb}. Among these approaches, the following interpretations are directly applicable to the formulation in this work:
\begin{align}
\tilde{X}_{\rm KH} &= \frac{1 + |X| - |Z|}{2},
\label{eq:X-KH}\\
\tilde{X} &= \frac{|X|}{|X| + |Z|}, 
\label{eq:X-tilde}\\
\mathcal{X} &= \frac{(\alpha_{0}-1)|X|-\alpha_{0}|Z|+\alpha_{0}}{2\alpha_{0}-1}, 
\label{eq:calX}
\end{align}
where $\alpha_{0}$ in the last formula is 
\begin{align}
\alpha_{0}&=\frac{\sqrt{5}-1+\sqrt{10-4\sqrt{5}}}{2}\approx 1.1318.
\end{align}
These interpretations were proposed in Refs.~\cite{Kamiya:2016oao}, \cite{Sekihara:2015gvw}, and \cite{Kinugawa:2024kwb}, respectively. In this way, $\tilde{X}_{\rm KH}$, $\tilde{X}$, and $\mathcal{X}$ can be regarded as probabilities because they take values between 0 and 1 under the corresponding conditions.

In addition, another prescription for the compositeness of resonances was proposed in Ref.~\cite{Matuschek:2020gqe}. Here, we extend this prescription to the case in the presence of the Coulomb interaction. In Ref.~\cite{Matuschek:2020gqe}, $\bar{X}_{A}$ is defined as the interpretable compositeness of resonances:
\begin{align}
\bar{X}_{A}(a_{s}, r_{e}) &= \sqrt{\frac{1}{1 + \left|\frac{2r_{e}}{a_{s}}\right|}},
\label{eq:XA-a-r}
\end{align}
by using the scattering length $a_{s}$ and the effective range $r_{e}$. The scattering length $a_{s}$ can be eliminated from Eq.~\eqref{eq:XA-a-r} using the bound-state condition for the short-range interaction:
\begin{align}
-\frac{1}{a_{s}} + \frac{r_{e}}{2}k_{h}^{2} - ik_{h} = 0.
\end{align}
As a result, $\bar{X}_{A}$ can be rewritten in terms of $r_{e}$ and the radius $R=i/k_{h}$ as
\begin{align}
\bar{X}_{A}(r_{e}, R) &= \sqrt{\frac{1}{1 + \left|-\left(\frac{r_{e}}{R}\right)^{2} + \frac{2r_{e}}{R}\right|}}.
\end{align}
We define $\bar{X}_{C}$ by replacing $r_{e}$ and $R$ with their Coulomb-modified counterparts $r_{e}^{C}$ and $R^{C}$:
\begin{align}
\bar{X}_{C}(r_{e}^{C}, R^{C}) &= \sqrt{\frac{1}{1 + \left|-\left(\frac{r_{e}^{C}}{R^{C}}\right)^{2} + \frac{2r_{e}^{C}}{R^{C}}\right|}}.
\label{eq:X-C}
\end{align}
We regard $\bar{X}_{C}$ as the compositeness of resonances in a system with Coulomb plus short-range interactions.

From Eq.~\eqref{eq:sigma-p}, $\bar{X}_{C}$ in Eq.~\eqref{eq:X-C} can also be written in terms of the energy derivative of the self-energy as
\begin{align}
\bar{X}_{C} &= \sqrt{\frac{1}{1 + \left|- \frac{1}{\Sigma'^{2}} + \frac{2}{\Sigma'}\right|}}.
\label{eq:XC-Sigma}
\end{align}
While Eq.~\eqref{eq:X-C} is equivalent to Eq.~\eqref{eq:XC-Sigma} in the present framework, we consider Eq.~\eqref{eq:XC-Sigma} to provide the compositeness of resonances for general EFTs. Although Eq.~\eqref{eq:XC-Sigma} requires the expression of the self-energy and is not always written in terms of the observables, it is applicable for arbitrary eigenenergies. On the other hand, the compositeness in Eq.~\eqref{eq:X-C} is written solely by the observables $r_{e}^{C}$ and $R^{C}$, but holds only for near-threshold states.

In the $|k_{h}| \to \infty$ limit with a fixed effective range, $R^{C} \to 0$. As a result, the complex compositeness $X$ tends to zero regardless of the direction from which $k_{h}$ approaches complex infinity. Therefore, even for resonances, the complex $X$ vanishes in this limit. Moreover, the interpretable compositeness of resonances introduced here tends to zero.

In the following, we employ the prescriptions in Eqs.~\eqref{eq:X-KH}, \eqref{eq:X-tilde}, \eqref{eq:calX}, and \eqref{eq:X-C} to interpret the complex compositeness of resonances.


\section{Properties of eigenstates: negative effective range}
\label{sec:negative-re}

In this section, we numerically investigate the properties of near-threshold eigenstates in systems with the Coulomb plus short-range interactions. We begin with the case of the negative effective range $r_{e}^{C}$ and compute the pole positions and the compositeness using the formulation introduced in the previous section. The results for repulsive Coulomb interactions are presented in Sec.~\ref{subsec:rep-negative-re}, followed by the attractive case in Sec.~\ref{subsec:att-negative-re}. The system with positive $r_{e}^{C}$ will be discussed in Sec.~\ref{sec:positive-re}.

\subsection{Repulsive Coulomb plus short-range interactions}
\label{subsec:rep-negative-re}

\subsubsection{Pole trajectory in momentum plane}

\begin{figure}[tbp]
\centering
\includegraphics[width=0.45\textwidth]{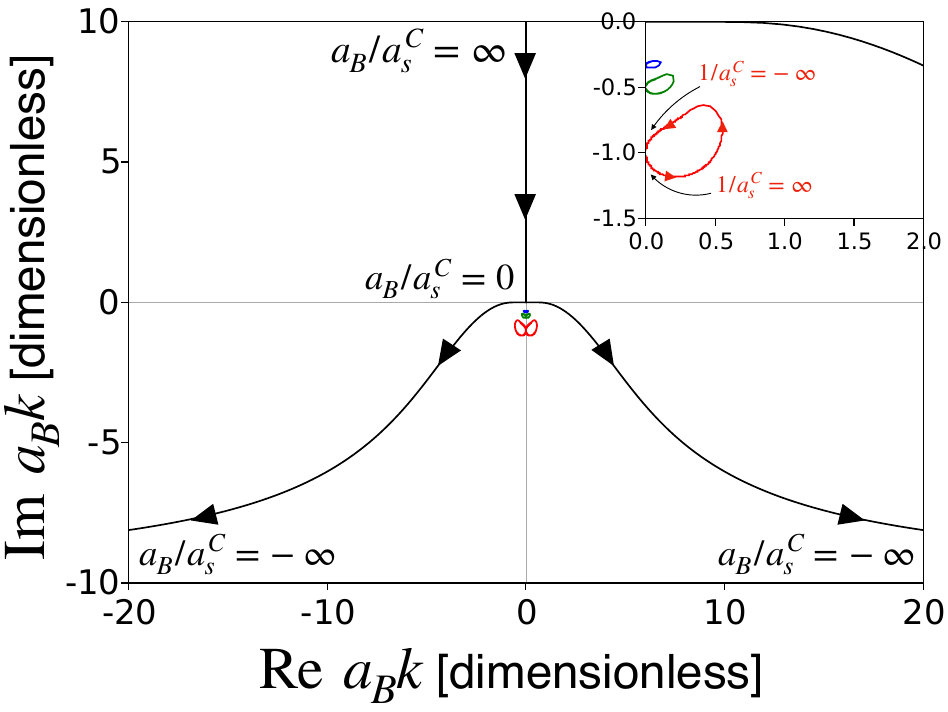}
\caption{Pole trajectories in the complex momentum plane for the repulsive Coulomb plus short-range system with $r_{e}^{C}/a_{B} = -0.1$ as the inverse scattering length $a_{B}/a_{s}^{C}$ is varied. For readability, we show the farthest three virtual poles from the threshold out of the infinitely many virtual poles. The arrows indicate the motion of the poles as $a_{B}/a_{s}^{C}$ decreases. The upper-right subfigure provides a close-up view of the near-threshold region along the imaginary axis.}
\label{fig:rep-neg-trajectory}
\end{figure}

In the framework in Sec.~\ref{sec:EFT}, the scattering amplitude can be determined by the three parameters: the scattering length $a_{s}^{C}$, the effective range $r_{e}^{C}$, and the Bohr radius $a_{B}$. To consider the negative effective range, $\sigma = + 1$ is adopted. The eigenmomentum is obtained by solving the pole condition~\eqref{eq:pole-condition-ERE} with given $a_{s}^{C}$, $r_{e}^{C}$, and $a_{B}$. We normalize all the quantities by $a_{B}$ and examine how the eigenmomentum changes as the scattering length $a_{s}^{C}$ is varied with a fixed $r_{e}^{C}$.

In Fig.~\ref{fig:rep-neg-trajectory}, we show the trajectory of the eigenmomentum, obtained by varying $a_{B}/a_{s}^{C}$ in the complex $a_{B}k$ plane. Throughout this plot, the effective range is fixed at $r_{e}^{C}/a_{B} = -0.1$. The arrows indicate the direction in which the pole moves as the inverse scattering length $a_{B}/a_{s}^{C}$ is changed from $-\infty$ to $\infty$. The origin $a_{B}k = 0$ corresponds to the scattering threshold. As discussed above, a logarithmic branch cut appears along the negative imaginary axis of the $a_{B}k$ plane.

Focusing on the region near the threshold along the imaginary axis in Fig.~\ref{fig:rep-neg-trajectory}, we observe the trajectories of the virtual-state poles, each of which has a finite width. Although an infinite number of virtual states exist, for clarity, we plot in Fig.~\ref{fig:rep-neg-trajectory} the three farthest trajectories from the threshold. To highlight their behavior, the virtual trajectories are shown in the upper-right inset in Fig.~\ref{fig:rep-neg-trajectory}. As the inverse scattering length $a_{B}/a_{s}^{C}$ is varied from $-\infty$ to $\infty$, each virtual pole traces a curved trajectory near the imaginary axis. The real part of the virtual pole vanishes in the $a_{B}/a_{s}^{C} \to \pm \infty$ limit. This behavior is consistent with Ref.~\cite{Mochizuki:2024dbf}, in which a zero-range theory expressed in terms of $a_{s}^{C}$ and $a_{B}$, corresponding to the further low-energy limit of the present model, is used to investigate the low-energy behavior of the scattering amplitude. In the $a_{B}/a_{s}^{C} \to \pm \infty$ limit, these virtual poles approach the pole in the system with the pure Coulomb interaction in Eq.~\eqref{eq:Coulomb-poles}. In this sense, these virtual poles are mainly formed by the Coulomb interaction. Because of this property, they are located near the threshold within the Coulomb scale $|a_{B}k| \lesssim 1$. 

Outside the Coulomb scale $|ka_{B}| \gtrsim 1$, we find a pole that originates from the pole formed by the short-range interaction. When $a_{B}/a_{s}^{C} \gg 1$, this pole lies on the positive imaginary $a_{B}k$ axis as a bound state. As $a_{B}/a_{s}^{C}$ decreases, the bound-state pole moves toward the threshold and its eigenmomentum becomes smaller. In the limit $a_{B}/a_{s}^{C} \to 0$, the pole reaches the threshold $a_{B}k = 0$. However, even at the threshold, the Bohr radius $a_{B}$ gives a finite scale, and therefore, there is no universal description of the system, in contrast to systems governed purely by short-range interactions~\cite{Mochizuki:2024dbf}.

When $a_{B}/a_{s}^{C}$ changes from positive to negative, the pole passes through the threshold and splits into two, becoming a resonance and an anti-resonance. It is important to note that the bound state directly turns into a resonance, although the eigenstate is in the $s$-wave scattering. In the system with the pure short-range interaction, an $s$-wave bound state turns into a virtual state when it crosses the threshold, while the direct transition from a bound state to a resonance occurs in partial waves with finite angular momentum $l \geq 1$~\cite{Taylor}. The similarity of the repulsive Coulomb potential and the infinitely strong centrifugal potential with $l \to \infty$ has been discussed in Ref.~\cite{Domcke:1983zz} on the basis of the Wigner threshold rule~\cite{Wigner:1948zz}. In fact, this bound-to-resonance transition is natural because the repulsive Coulomb interaction plays the role of a barrier, similarly to the centrifugal potential with the finite angular momentum at a sufficiently large distance, where the short-range interaction is negligible.

There is, however, an important difference in the bound-to-resonance transition with and without the Coulomb interaction from the viewpoint of pole-number conservation. In the absence of the Coulomb interaction, when a bound-state pole approaches the threshold, its accompanying virtual-state pole also moves toward the threshold. The two poles then meet at the threshold and subsequently evolve into a resonance and an anti-resonance. Therefore, the number of poles along the trajectory is conserved. In general, the argument principle ensures that the number of poles inside a closed contour in the complex plane remains conserved under the variation of parameters~\cite{Kamiya:2017pcq}. 

In contrast, in the presence of the Coulomb interaction, this number conservation appears not to hold at the threshold since one bound state turns into a resonance and an anti-resonance in Fig.~\ref{fig:rep-neg-trajectory}. This apparent splitting of a bound-state pole into a resonance and an anti-resonance at the threshold may not contradict the pole-number conservation, because the logarithmic branch cut prevents one from enclosing the threshold by a closed contour. A similar issue arises in Efimov physics, where the number of poles does not seem to be conserved when a three-body bound state turns into a resonance at the threshold~\cite{Hyodo:2013zxa,Dawid:2023kxu}.

\begin{figure}[tbp]
\centering
\includegraphics[width=0.45\textwidth]{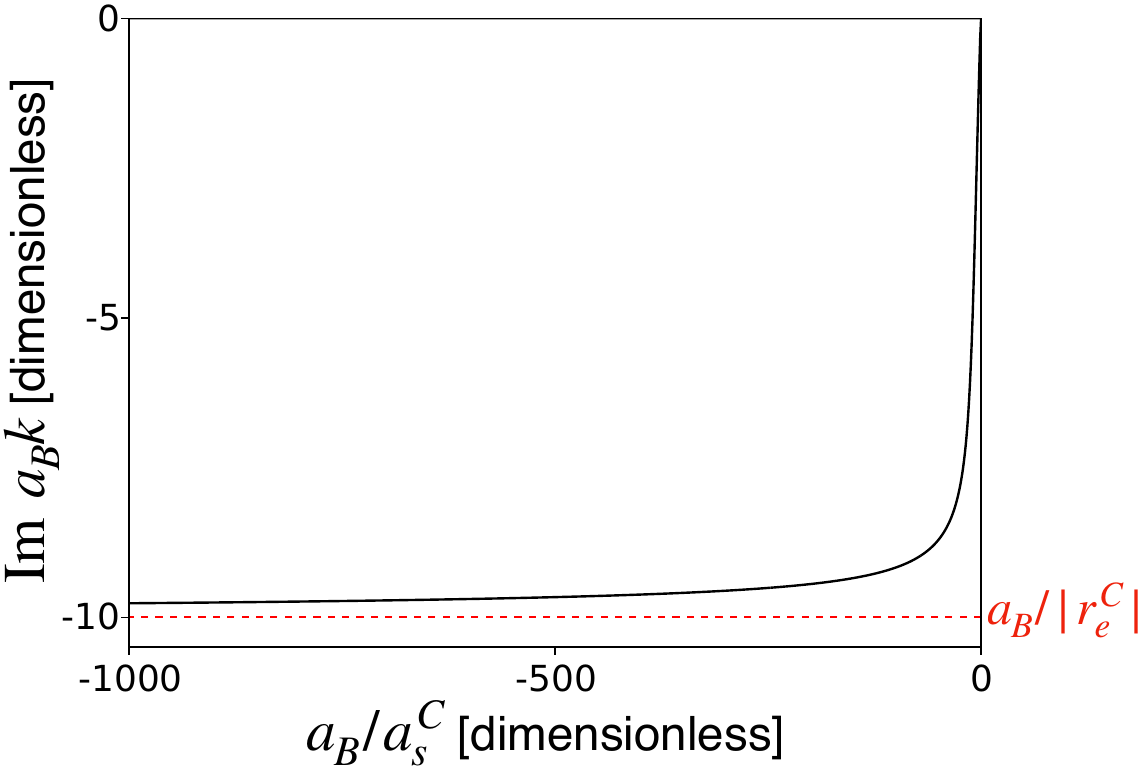}
\caption{Imaginary part of the resonance eigenmomentum $\Im\, a_{B}k$ in the repulsive Coulomb plus short-range system with $r_{e}^{C}/a_{B} = -0.1$. The dashed line indicates $\Im\, a_{B}k = a_{B}/|r_{e}^{C}|$.}
\label{fig:rep-neg-large-k}
\end{figure}

As $a_{B}/a_{s}^{C}$ becomes negatively large, the resonance and anti-resonance move away from the threshold, and their imaginary parts appear to approach a constant value. To clarify this behavior in the region far from the threshold, we examine the resonance pole in the large-$k$ limit, $|a_{B}k|\to\infty$, while keeping $r_{e}^{C}$ fixed. In Fig.~\ref{fig:rep-neg-large-k}, we plot the imaginary part of the eigenmomentum of the resonance as a function of the inverse scattering length $a_{B}/a_{s}^{C}$. The figure shows that the imaginary part asymptotically approaches $a_{B}/|r_{e}^{C}|$. This behavior can be understood analytically in Eq.~\eqref{eq:k-pm}, where the pole trajectory asymptotically approaches that of the pure short-range interaction.

\begin{figure*}[tbp]
\centering
\includegraphics[width=0.45\textwidth]{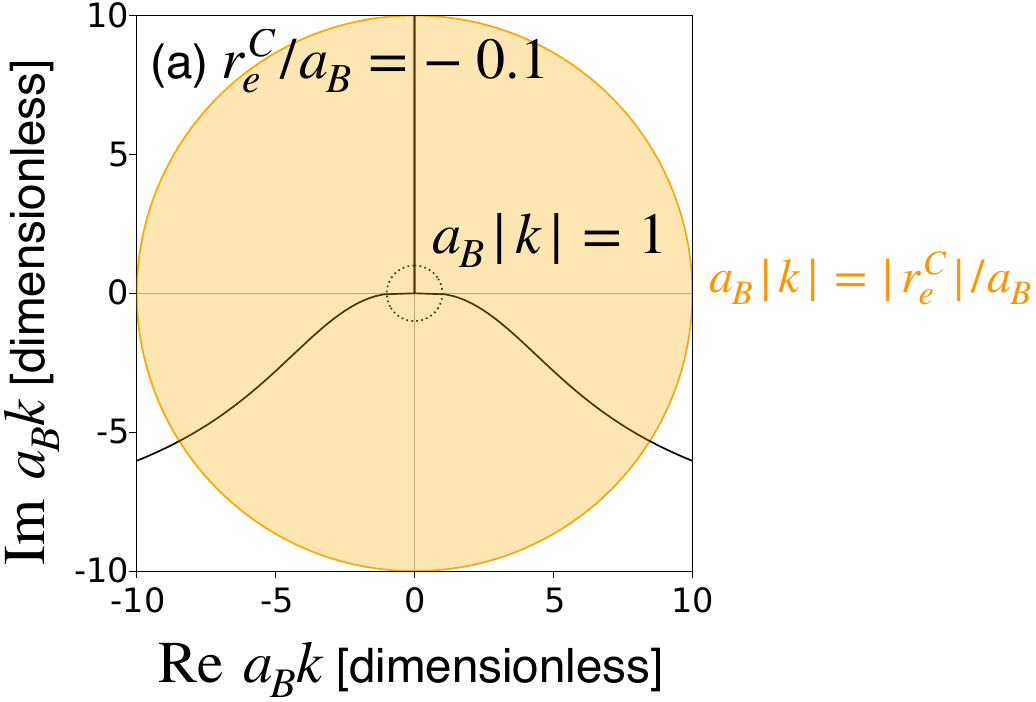}
\includegraphics[width=0.32\textwidth]{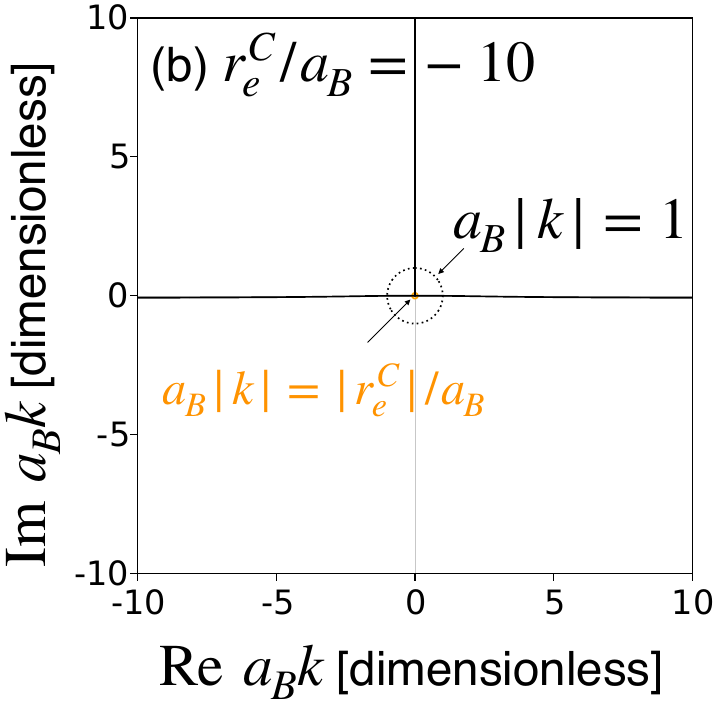}
\caption{Pole trajectories in the complex momentum plane for the repulsive Coulomb plus short-range system with $r_{e}^{C}/a_{B} = -0.1$ [panel (a)] and $r_{e}^{C}/a_{B} = -10$ [panel (b)]. The dotted lines indicate the boundary $|ka_{B}| = 1$, while the orange solid lines indicate the boundary $|ka_{B}| = a_{B}/|r_{e}^{C}|$. }
\label{fig:rep-neg-re}
\end{figure*}

So far, the effective range $r_{e}^{C}/a_{B}$ has been fixed at $-0.1$. We now examine how the pole trajectories change when $r_{e}^{C}/a_{B}$ is varied. In Fig.~\ref{fig:rep-neg-re}, we show the trajectories of bound states, resonances, and anti-resonances for $r_{e}^{C}/a_{B} = -0.1$ [panel (a)] and for $r_{e}^{C}/a_{B} = -10$ [panel (b)]. The trajectories of the virtual states are omitted for clarity. By comparing with panel (a) ($r_{e}^{C}/a_{B} = -0.1$), the width of the resonance is small in panel (b) ($r_{e}^{C}/a_{B} = -10$). From Eq.~\eqref{eq:re-C}, a large effective range is obtained for a small coupling constant. In this case, the resonance does not couple to the scattering states very much, which leads to the small decay width. In this sense, it is natural to obtain the narrow resonance with the large effective range $r_{e}^{C}/a_{B} = -10$.

To visualize the competition between the Coulomb and short-range interactions, we draw the boundaries $|a_{B}k| = 1$ (dotted lines) and $|a_{B}k| = a_{B}/|r_{e}^{C}|$ (solid lines) in Fig.~\ref{fig:rep-neg-re}, which represent the characteristic momentum scales of the Coulomb and short-range interactions, respectively. When $1/|a_{s}^{C}| \ll 1/|r_{e}^{C}|$, the pole position can be written solely by the scattering length because the $r_{e}^{C}k^{2}/2$ term can be neglected in the scattering amplitude~\eqref{eq:fCS-renorm}. In other words, when the pole is located far inside the solid circle, its position can be determined by $a_{s}^{C}$ as in the zero-range theory in Ref.~\cite{Mochizuki:2024dbf}. On the other hand, when $1/|a_{s}^{C}| \sim 1/|a_{B}|$, namely, the pole exists near the dotted circle, the Coulomb interaction is not negligible to determine the pole position. Therefore, if there exists a region sufficiently close to the threshold that the contribution of $r_{e}^{C}$ to the pole position is negligible, while still being sufficiently far from the threshold that the Coulomb interaction does not contribute, the pole position can be universally determined by the scattering length. For such a region to exist, the condition $1/|a_{B}| \ll 1/|a_{s}^{C}| \ll 1/|r_{e}^{C}|$ is required. From the viewpoint of the parameters of the system, a universal region emerges for $|r_{e}^{C}|/a_{B} < 1$, which, as discussed in Table~\ref{tab:limits-sum}, corresponds to a system with a weak Coulomb interaction. A representative example is panel~(a) of Fig.~\ref{fig:rep-neg-re}, where $r_{e}^{C}/a_{B} = -0.1$. Indeed, as shown in the plot, there is a universal region indicated by the shaded area in panel~(a).

When $a_{B}|k| < 1$, the pole approaches the region where the eigenstates generated mainly by the Coulomb interaction are originally located. In this regime, the Coulomb and short-range interactions become comparable in strength, and the Coulomb contribution cannot be neglected. Consequently, both $a_{s}^{C}$ and $a_{B}$ become relevant length scales in the pole condition~\eqref{eq:pole-condition-ERE}, and the low-energy universality is not realized.

In panel~(b) with $r_{e}^{C}/a_{B} = -10$, as $a_{B}|k|$ decreases, the Coulomb-dominant region $a_{B}|k| < 1$ appears before the region in which the short-range universality would be expected. Therefore, throughout the region with $a_{B}|k| < a_{B}/|r_{e}^{C}|$, the contribution of the Coulomb interaction becomes important. As a result, there is no region in which the universality appears in this setup.

\begin{figure*}[tbp]
\centering
\includegraphics[width=0.9\textwidth]{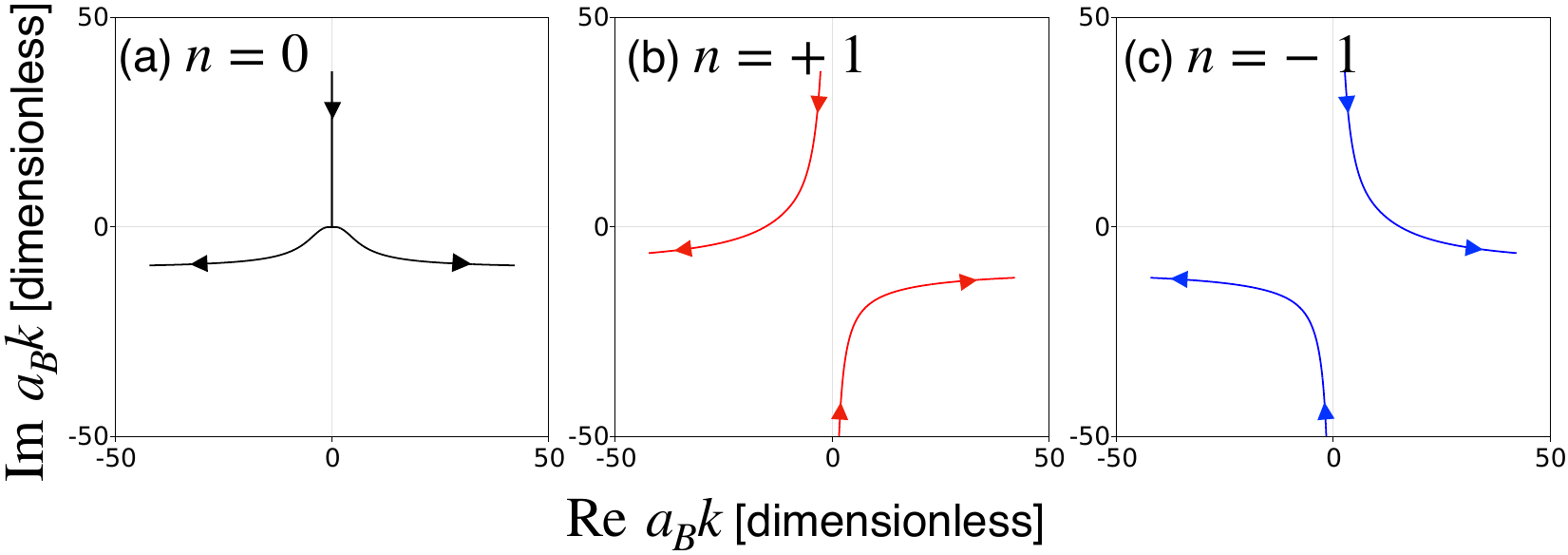}
\caption{Pole trajectories in different Riemann sheets; (a) in the $n = 0$ sheet (same as Fig.~\ref{fig:rep-neg-trajectory}), (b) in the $n = +1$ sheet, and (c) in the $n = -1$ sheet. The near-threshold virtual poles are not shown here.}
\label{fig:rep-neg-Riemann}
\end{figure*}

In addition to the $n = 0$ sheet, which has been discussed so far, we now turn to other Riemann sheets. In Fig.~\ref{fig:rep-neg-Riemann}, we show the pole trajectories on the $n = 0$ sheet [panel (a)], the $n = +1$ sheet [panel (b)], and the $n = -1$ sheet [panel (c)] with the fixed effective range $r_{e}^{C}/a_{B} = -0.1$. In Fig.~\ref{fig:rep-neg-Riemann}, the Coulomb-originated virtual states are not shown. The arrows indicate the motion of the pole as the inverse scattering length is varied from $a_{B}/a_{s}^{C} = \infty$ to $a_{B}/a_{s}^{C} = -\infty$.

In panel (a) (the $n = 0$ sheet), the pole trajectory is symmetric with respect to the imaginary axis. In contrast, in panels (b) and (c), the trajectories do not exhibit symmetry. Comparing panels~(b) and~(c), the pole trajectories exhibit mutually symmetric behavior. In general, the Schwarz reflection principle indicates that the pole trajectory in the $n = 0$ sheet is symmetric within a single sheet, while the $n = +m$ and $n = -m$ sheets form mirror-image pairs~\cite{Dawid:2023kxu}.

In the purely short-range system, if there is a deeply bound state originating from the bare state, a corresponding deep virtual state also appears~\cite{Morgan:1992ge,Baru:2003qq,Hyodo:2014bda,Hanhart:2022qxq}. However, in the presence of the Coulomb interaction, no virtual-state pole appears on the negative imaginary axis because of the branch cut. The virtual poles located near the imaginary axis in the $n = \pm 1$ sheet can be interpreted as originating from the virtual state paired with the bound state.

In the region far from the origin, poles exist on the $n = \pm 1$ sheets near the bound-state, resonance, and anti-resonance poles on the $n = 0$ sheet. The motions of these poles on the $n = \pm 1$ sheet are correlated with those of the corresponding poles on the $n = 0$ sheet. Thus, the poles on the $n = \pm 1$ sheets can be regarded as shadow poles of the corresponding poles on the $n=0$ sheet.

\subsubsection{Eigenenergy}

\begin{figure*}[tbp]
\centering
\includegraphics[width=0.45\textwidth]{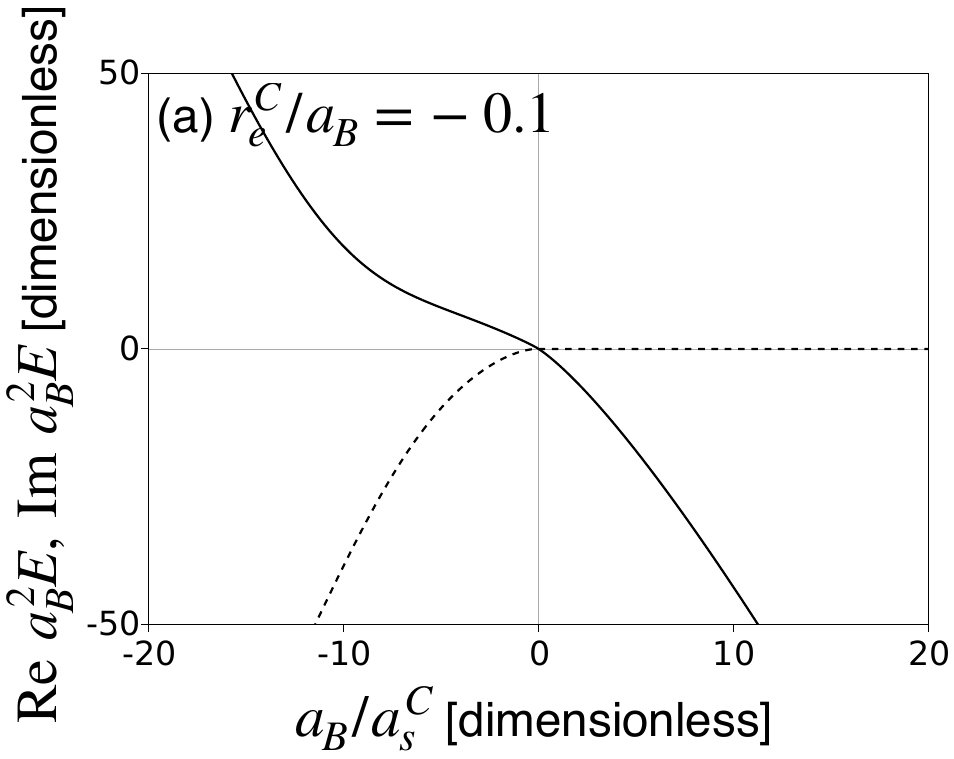}
\includegraphics[width=0.45\textwidth]{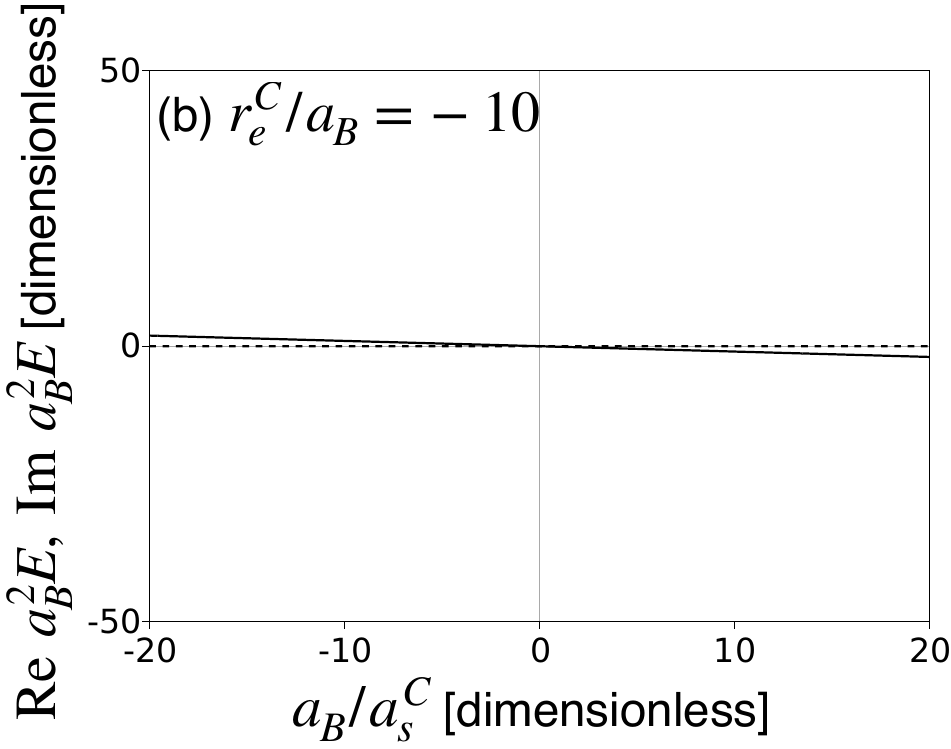}
\caption{The real and imaginary parts of the eigenenergy as functions of the inverse scattering length $a_{B}/a_{s}^{C}$ in the repulsive Coulomb plus short-range system with $r_{e}^{C}/a_{B} = -0.1$ [panel (a)] and $r_{e}^{C}/a_{B} = -10$ [panel (b)]. The solid lines represent the real part of the eigenenergy, and the dashed lines represent the imaginary part.}
\label{fig:rep-neg-E}
\end{figure*}

To further investigate the nature of near-threshold bound states and resonances, in Fig.~\ref{fig:rep-neg-E} we plot the real part (solid lines) and imaginary part (dashed lines) of the eigenenergy as functions of $a_{B}/a_{s}^{C}$ for the pole that evolves from a bound state into a resonance. Panel (a) shows the case with $r_{e}^{C}/a_{B} = -0.1$, and panel (b) corresponds to $r_{e}^{C}/a_{B} = -10$. In the region $a_{B}/a_{s}^{C} > 0$, the imaginary part of the eigenenergy is exactly zero because the pole represents a bound state. For $a_{B}/a_{s}^{C} < 0$ i.e., beyond the unitary limit, an imaginary part appears, indicating that the bound state turns into a resonance after crossing the threshold, as shown in Fig.~\ref{fig:rep-neg-trajectory}. This qualitative behavior is similar to that in higher partial waves discussed in Ref.~\cite{Hyodo:2014bda}. By focusing on the imaginary part of the eigenenergy, the resonance in panel (a) has a broader width than that in panel (b). This is consistent with the discussion around Fig.~\ref{fig:rep-neg-re} that a narrow resonance appears in the system with a large $|r_{e}^{C}|/a_{B}$. 

From the slope of $\Re\, E$ at $a_{B}/a_{s}^{C} = 0$, we can infer the compositeness of the bound state at the threshold. In the absence of the Coulomb interaction, the low-energy universality indicates that the real part of the eigenenergy shows a quadratic dependence on the inverse scattering length, Re~$E \sim (a_{B}/a_{s}^{C})^{2}$. In this case, the compositeness becomes unity, as reflected in the vanishing slope of $\Re\, E$ at the threshold~\cite{Hyodo:2014bda}. In contrast, in the presence of the Coulomb interaction, the slope of Re~$E$ at the threshold is finite in both panels (a) and (b), indicating that $X \neq 1$. 

\subsubsection{Compositeness}

\begin{figure*}[tbp]
\centering
\includegraphics[width=0.45\textwidth]{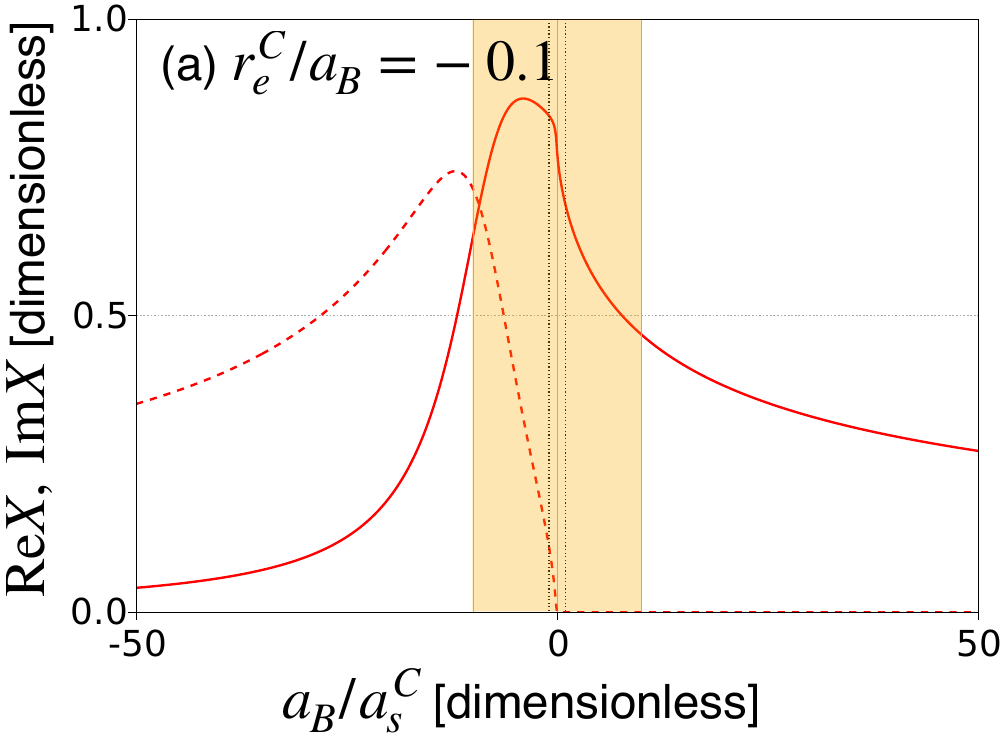}
\includegraphics[width=0.45\textwidth]{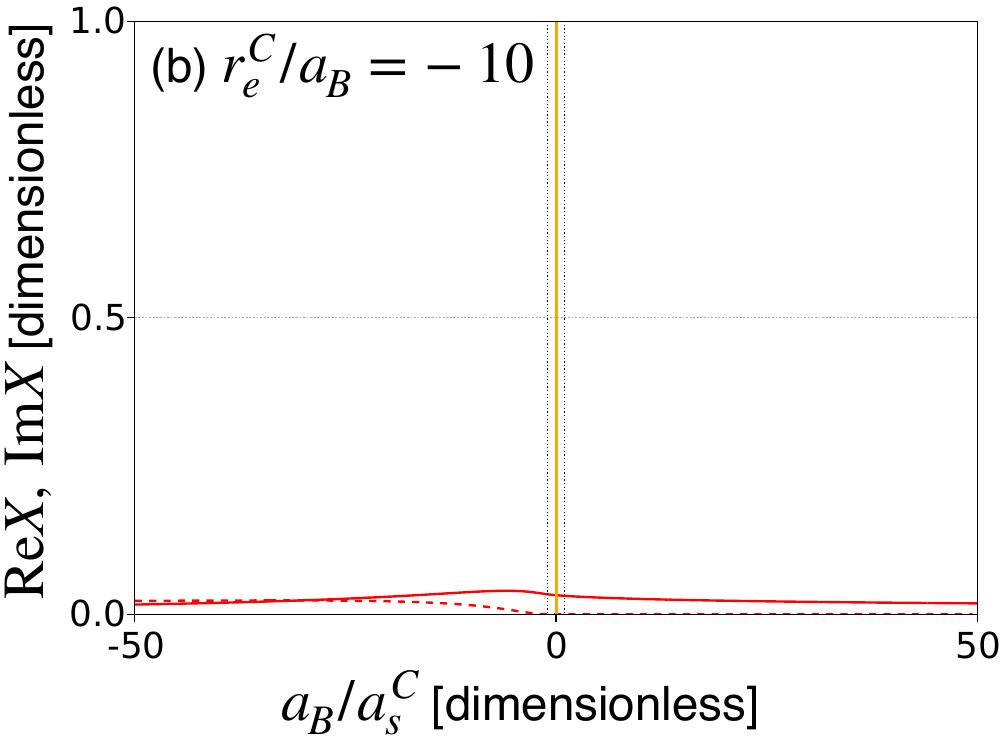}
\caption{The real and imaginary parts of the compositeness $X$ as functions of the inverse scattering length $a_{B}/a_{s}^{C}$ in the repulsive Coulomb plus short-range system with $r_{e}^{C}/a_{B} = -0.1$ [panel (a)] and $r_{e}^{C}/a_{B} = -10$ [panel (b)]. The solid lines represent the real part of $X$, and the dashed lines represent the imaginary part.}
\label{fig:rep-neg-X}
\end{figure*}

To quantitatively investigate the internal structure of near-threshold states, Fig.~\ref{fig:rep-neg-X} shows the real parts (solid lines) and imaginary parts (dashed lines) of the compositeness $X$ as functions of the inverse scattering length $a_{B}/a_{s}^{C}$ for the pole that evolves from a bound state to a resonance. Similar to Fig.~\ref{fig:rep-neg-E}, panel (a) corresponds to the case with $r_{e}^{C}/a_{B} = -0.1$, and panel (b) to the case with $r_{e}^{C}/a_{B} = -10$. As a guideline for identifying relevant interactions, we also indicate $a_{B}/a_{s}^{C} = \pm 1$ with vertical dotted lines and $a_{B}/a_{s}^{C} = \pm a_{B}/|r_{e}^{C}|$ with vertical solid lines.

In the bound-state region $a_{B}/a_{s}^{C} > 0$, the compositeness is real by definition. In contrast, in the resonance region $a_{B}/a_{s}^{C} < 0$, $X$ becomes complex and develops a finite imaginary part, as discussed above. In both panels (a) and (b), bound states located far from the threshold (large $|k_{h}|$) with large values of $a_{B}/a_{s}^{C}$ exhibit small compositeness. This behavior follows from the analytic property that $X \to 0$ in the limit $|k_{h}| \to \infty$ as discussed in Sec.~\ref{subsec:expression-X}.

In contrast, the behavior near $a_{B}/a_{s}^{C} = 0$ differs clearly between panels (a) and (b). This difference in the near-threshold compositeness is suggested by the discussion around Fig.~\ref{fig:rep-neg-re}: in panel (a) with $r_{e}^{C}/a_{B} = -0.1$, remnants of the short-range universality are expected to appear, whereas in panel (b) with $r_{e}^{C}/a_{B} = -10$, the Coulomb interaction becomes important before the short-range universality can emerge. As a result, in panel (a), the presence of a short-range universal regime $1 \ll a_{B}/a_{s}^{C} \ll a_{B}/|r_{e}^{C}|$ leads to an increase in $X$. However, very close to the threshold $a_{B}/a_{s}^{C} \lesssim 1$, the Coulomb interaction is not negligible. Consequently, $X$ stays large but does not reach unity at the threshold. In contrast, in panel (b), no enhancement of $X$ is observed near the threshold because there is no enhancement mechanism for $X$ in this region. 

\begin{figure*}
\centering
\includegraphics[width=0.45\textwidth]{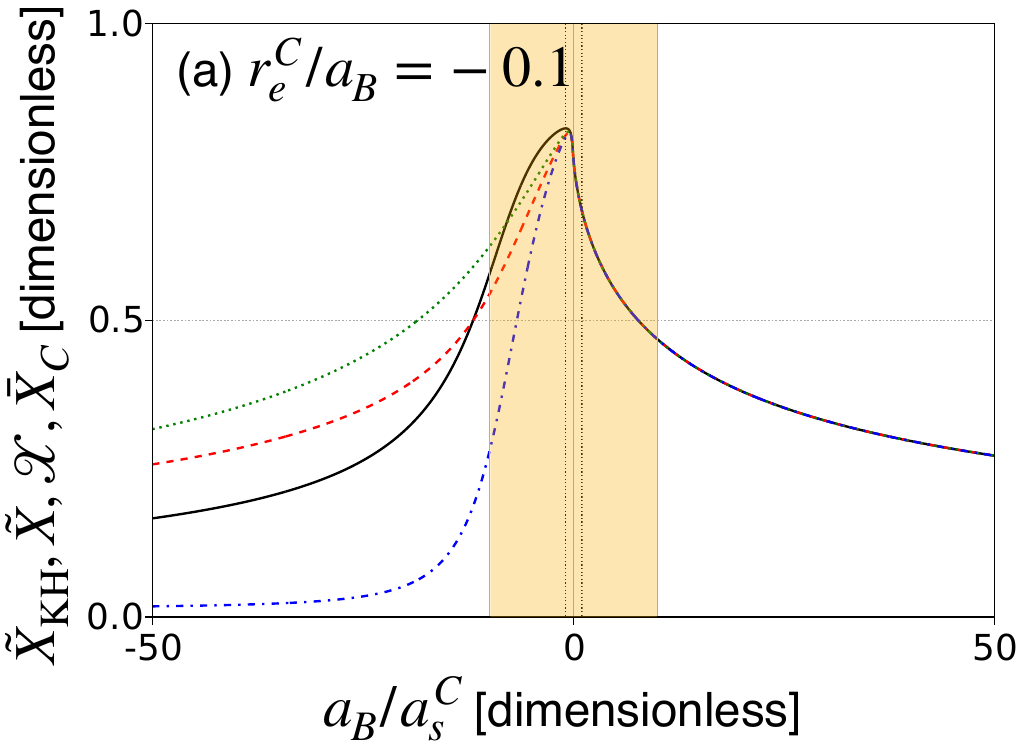}
\includegraphics[width=0.45\textwidth]{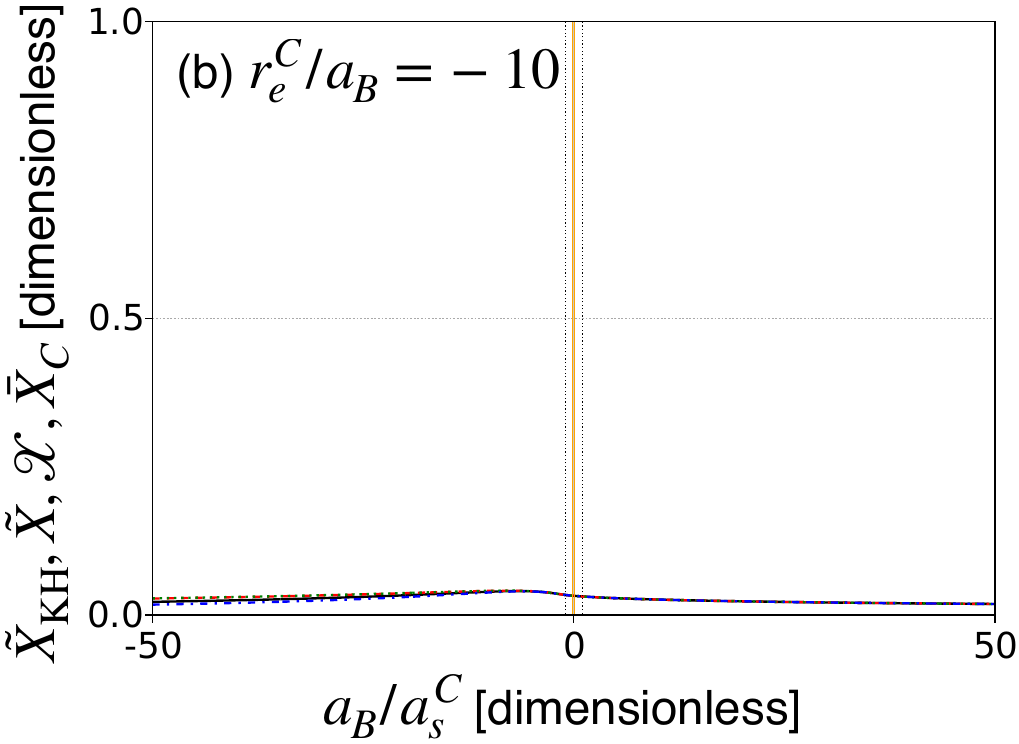}
\caption{Compositeness in the interpretation schemes introduced in Sec.~\ref{subsec:compositeness} as functions of the inverse scattering length $a_{B}/a_{s}^{C}$ in the repulsive Coulomb plus short-range system with $r_{e}^{C}/a_{B} = -0.1$ [panel (a)] and $r_{e}^{C}/a_{B} = -10$ [panel (b)]. The solid, dashed, dash-dotted, and dotted lines represent $\tilde{X}_{\rm KH}$, $\tilde{X}$, $\mathcal{X}$, and $\bar{X}_{C}$, respectively.}
\label{fig:rep-neg-interpretations}
\end{figure*}

To quantify the internal structure of resonances in the region with negative scattering length $a_{B}/a_{s}^{C} < 0$, we apply the interpretation schemes introduced in Sec.~\ref{subsec:compositeness}. Figure~\ref{fig:rep-neg-interpretations} shows $\tilde{X}_{\rm KH}$ (solid line), $\tilde{X}$ (dashed line), $\mathcal{X}$ (dash-dotted line), and $\bar{X}_{C}$ (dotted line). Panels (a) and (b) correspond to the systems with $r_{e}^{C}/a_{B} = -0.1$ and $r_{e}^{C}/a_{B} = -10$, respectively. Again, we show $a_{B}/a_{s}^{C} = \pm 1$ as vertical dotted lines and $a_{B}/a_{s}^{C} = \pm a_{B}/|r_{e}^{C}|$ as vertical solid lines.

By construction, all interpretation schemes coincide in the bound-state region $a_{B}/a_{s}^{C} > 0$, where the compositeness is real. In contrast, in the resonance region with the negative scattering length $a_{B}/a_{s}^{C} < 0$, the results deviate from each other. In the near-threshold region, quantitative differences among the schemes remain small, while they become more pronounced away from the threshold. As shown in Fig.~\ref{fig:rep-neg-X}, this behavior originates from the growth of the imaginary part of the complex compositeness in the far-threshold region, which increases the ambiguity in the interpretation. Nevertheless, the qualitative conclusion regarding whether a resonance is composite dominant or elementary dominant remains essentially unchanged.

In panel (a), owing to the remnant of the short-range universality, the compositeness remains close to unity even for near-threshold resonances. Together with the bound-state results, this indicates that the states are composite dominant within the short-range universal region, roughly given by $-a_{B}/|r_{e}^{C}| \lesssim a_{B}/a_{s}^{C} \lesssim a_{B}/|r_{e}^{C}|$. For resonances located far from the threshold with negatively large $a_{B}/a_{s}^{C}$, the results of the compositeness become small, indicating that these states are elementary dominant. This behavior reflects the analytic properties of the compositeness for large $|k_{h}|$. In contrast, in panel (b), as suggested by the bound-state results, even near-threshold resonances remain elementary dominant, characterized by the small compositeness.

Furthermore, the compositeness varies continuously even when a bound state turns into a resonance at $a_{B}/a_{s}^{C} = 0$. It is natural that the internal structure is smoothly connected, since the pole moves continuously with respect to $a_{B}/a_{s}^{C}$. From panel (a) of Fig.~\ref{fig:rep-neg-interpretations}, we find that a near-threshold resonance is composite dominant if a shallow bound state is composite dominant because of the remnant of the short-range universality. This is in sharp contrast to systems with purely short-range interactions, in which the internal structures of near-threshold resonances are necessarily non-composite dominant~\cite{Hyodo:2013iga,Matuschek:2020gqe,Hanhart:2022qxq,Kinugawa:2023fbf,Kinugawa:2024kwb}. In this way, the internal structure of near-threshold narrow resonances can drastically change with or without the Coulomb interaction. 

Recall that the system with the weak Coulomb interaction $|r_{e}^{C}|/a_{B} < 1$ is shown in panel (a), where the resonance structure drastically changes by switching off the Coulomb interaction. However, the composite-dominant resonance in panel (a) appears in the near-threshold region with $-a_{B}/|r_{e}^{C}| < a_{B}/a_{s}^{C} < 0$. In this region, in the absence of the Coulomb interaction, a virtual state appears instead of a resonance. Conversely, in the far-threshold region $a_{B}/a_{s}^{C} < -a_{B}/|r_{e}^{C}|$, where a resonance emerges without the Coulomb interaction, the resonance with the Coulomb interaction is non-composite dominant in panel (a), which is consistent with the result for the short-range interaction system.

We have seen that in the presence of the repulsive Coulomb interaction, a bound state directly turns into a resonance without being a virtual state. We summarize the important findings in this section:
\begin{itemize}
\item When the Coulomb interaction is weak ($|r_{e}^{C}|/a_{B} < 1$), the remnant of the short-range universality appears, and the near-threshold eigenstates are composite dominant.
\item When the Coulomb interaction is strong ($|r_{e}^{C}|/a_{B} > 1$), there are no effects that enhance the compositeness, and the eigenstates can be elementary-dominant even in the near-threshold region.
\end{itemize}
We show the universal nature of near-threshold eigenstates in Table~\ref{tab:X-sum}.

\begin{table}
 \caption{Universal properties of near-threshold eigenstates.\label{tab:X-sum}}
  \begin{ruledtabular}
  \begin{tabular}{lccc}
    Eigenstates & $|r_{e}^{C}|/a_{B} > 1$ & $|r_{e}^{C}|/a_{B} < 1$ & $|r_{e}^{C}|/a_{B} = 0$ \\ \hline 
    Bound states & - & composite & composite \\
    Resonances & - & composite & elementary \\
  \end{tabular}
  \end{ruledtabular}
\end{table}

\subsection{Attractive Coulomb plus short-range interactions}
\label{subsec:att-negative-re}

\subsubsection{Pole trajectory in momentum plane}

\begin{figure}[tbp]
\centering
\includegraphics[width=0.45\textwidth]{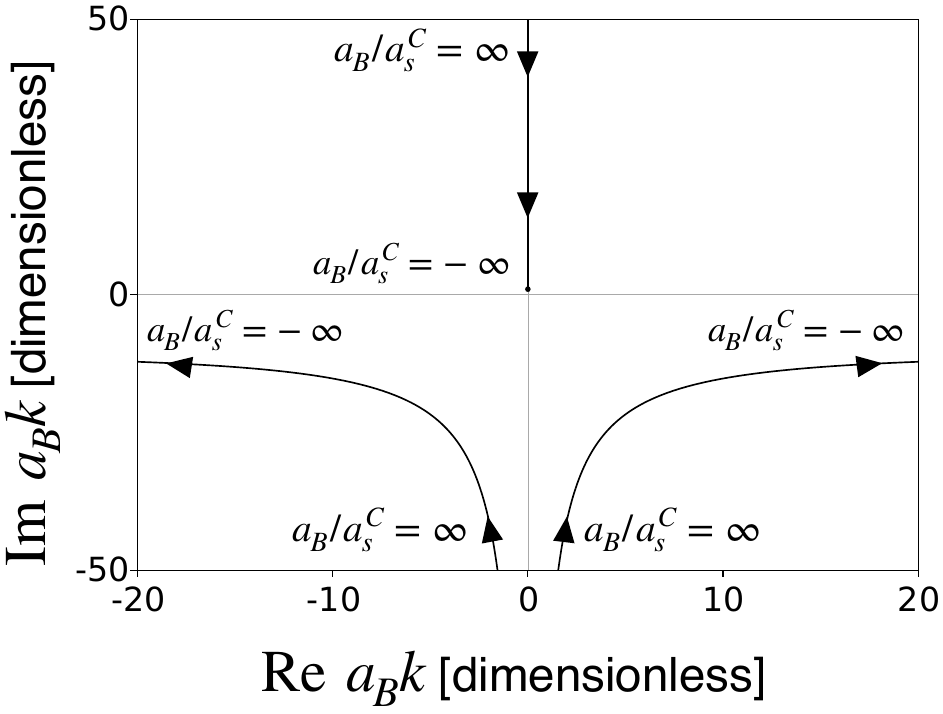}
\caption{Pole trajectories in the complex momentum plane for the attractive Coulomb plus short-range system with $r_{e}^{C}/a_{B} = -0.1$ as the inverse scattering length $a_{B}/a_{s}^{C}$ is varied. The arrows indicate the motion of the poles as $a_{B}/a_{s}^{C}$ decreases. The near-threshold bound poles are not shown here.}
\label{fig:att-neg-large-k}
\end{figure}

From now on, we consider the attractive Coulomb interaction on top of the short-range interaction. As with the repulsive case, we begin by examining the pole trajectory in the complex $a_{B}k$ plane. We first consider bound states on the positive imaginary axis. When the inverse scattering length $a_{B}/a_{s}^{C}$ is sufficiently large, there are a deep bound state ($a_{B}|k| \gg 1$) formed mainly by the short-range interaction and infinitely many shallow bound states ($a_{B}|k| \lesssim 1$) generated by the attractive Coulomb interaction, as discussed in Sec.~\ref{subsec:poles}.

We focus on the bound state originating from the short-range interaction, which exists far from the threshold when $a_{B}/a_{s}^{C}$ is positive and sufficiently large. The bound state becomes shallower with decreasing $a_{B}/a_{s}^{C}$, but it remains a bound state even when the sign of $a_{B}/a_{s}^{C}$ is flipped. In the $1/a_{s}^{C}\to -\infty$ limit, the pole approaches the Coulomb ground state level at $k = i/a_{B}$ in accordance with the analysis in Sec.~\ref{subsec:poles}. This means that a bound state originating from the short-range interaction does not turn into a virtual state or a resonance for any parameter choice. 

In addition to the bound state discussed above, there exist infinitely many bound states near the threshold generated by the Coulomb interaction. In the $a_{B}/a_{s}^{C} \to \infty$ limit, these bound states are located at the Coulomb levels $k = i/(a_{B}n)$. As $a_{B}/a_{s}^{C}$ decreases, these states become gradually shallower, and in the $a_{B}/a_{s}^{C} \to -\infty$ limit, each state shifts to the next shallower Coulomb level relative to its original one~\cite{Domcke:1983zz,Mochizuki:2024dbf}. In other words, the state initially at the $n=1$ Coulomb level moves to the $n=2$ level, the state at $n = 2$ moves to $n = 3$, and so on. This behavior can be understood as a consequence of level repulsion due to the approach of the bound state originating from the short-range interaction.

We then focus on the lower half of the $a_{B}k$ plane. When $a_{B}/a_{s}^{C}$ is sufficiently large, a pair of virtual states exists at $a_{B}|k| \gg 1$ near the negative imaginary axis. As the inverse scattering length decreases, the virtual states approach the threshold and evolve into a resonance and an anti-resonance at around $a_{B}/a_{s}^{C} = 0$. In the limit of $a_{B}/a_{s}^{C} \to -\infty$, the pole trajectories of the resonance and anti-resonance asymptotically approach $\Im\, a_{B}k = -a_{B}/|r_e^C|$ as in the short-range interaction system. This behavior is also seen in the repulsive Coulomb case.

In Fig.~\ref{fig:att-neg-large-k}, we plot the pole trajectories except for the Coulomb bound states. We see that the trajectory of the bound-state pole along the positive imaginary axis is not continuously connected to the resonance trajectory in the lower half of the $a_{B}k$ plane. In the presence of the attractive Coulomb interaction, the pole trajectory is considered to be discontinuous because infinitely many Coulomb bound-state poles accumulate near the origin.

\subsubsection{Compositeness}
\label{subsubsec:X-att-neg}

\begin{figure}[tbp]
\centering
\includegraphics[width=0.45\textwidth]{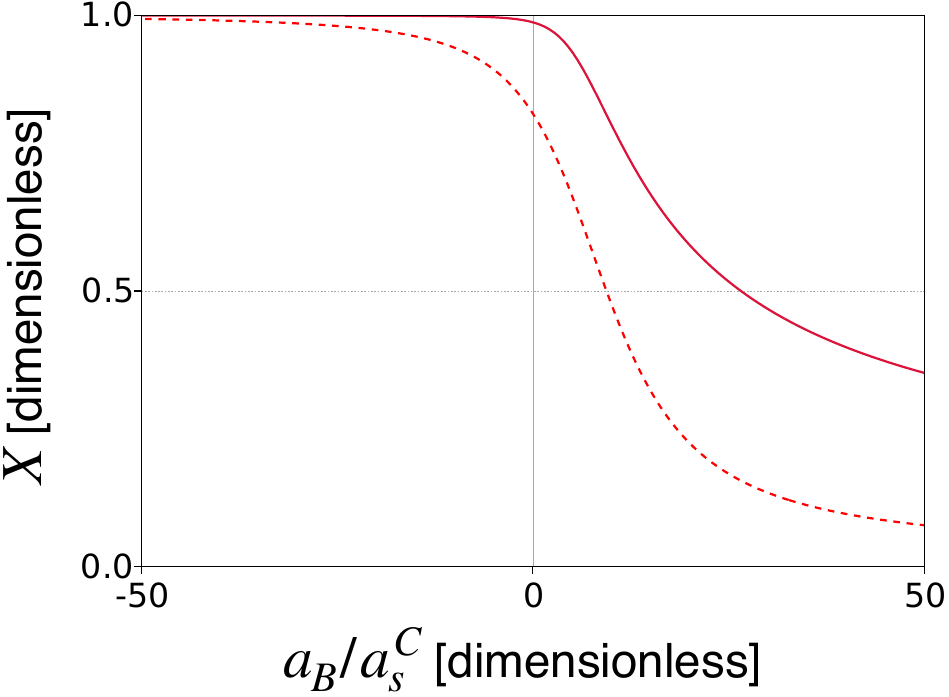}
\caption{Compositeness $X$ as functions of the inverse of the scattering length $a_{B}/a_{s}^{C}$. The solid (dashed) line represents the result with $r_{e}^{C}/a_{B} = -0.1$ ($r_{e}^{C}/a_{B} = -10$). }
\label{fig:att-neg-X}
\end{figure}

We now evaluate the compositeness of eigenstates in the system with the attractive Coulomb plus short-range interactions. We first focus on the bound state originating from the short-range interaction. Figure~\ref{fig:att-neg-X} shows the $a_{B}/a_{s}^{C}$ dependence of the compositeness of the bound state with the effective range fixed at $r_{e}^{C}/a_{B} = -0.1$ (solid line) and $r_{e}^{C}/a_{B} = -10$ (dashed line). In both cases, the compositeness $X$ is small when $a_{B}/a_{s}^{C}$ is large. This is similar to the repulsive case and can be attributed to the fact that this bound state is generated from the bare state with $X = 0$ in the model. As $a_{B}/a_{s}^{C}$ decreases, the bound state becomes shallower and the compositeness increases. In particular, $X$ increases rapidly when the sign of $a_{s}^{C}$ changes. In the limit $a_{B}/a_{s}^{C} \to -\infty$, $X$ approaches unity regardless of the value of the effective range. This is because in this limit, the bound state becomes the Coulomb ground state with $X = 1$, which contains no bare-state component.
Focusing on the $r_{e}^{C}$ dependence of the compositeness, we find that $X$ approaches unity more rapidly for a small effective range ($|r_{e}^{C}|/a_{B} = 0.1$) with the weak Coulomb interaction. Similar to the repulsive Coulomb interaction, this behavior can be explained by a remnant of short-range universality. 

\begin{figure}[tbp]
\centering
\includegraphics[width=0.45\textwidth]{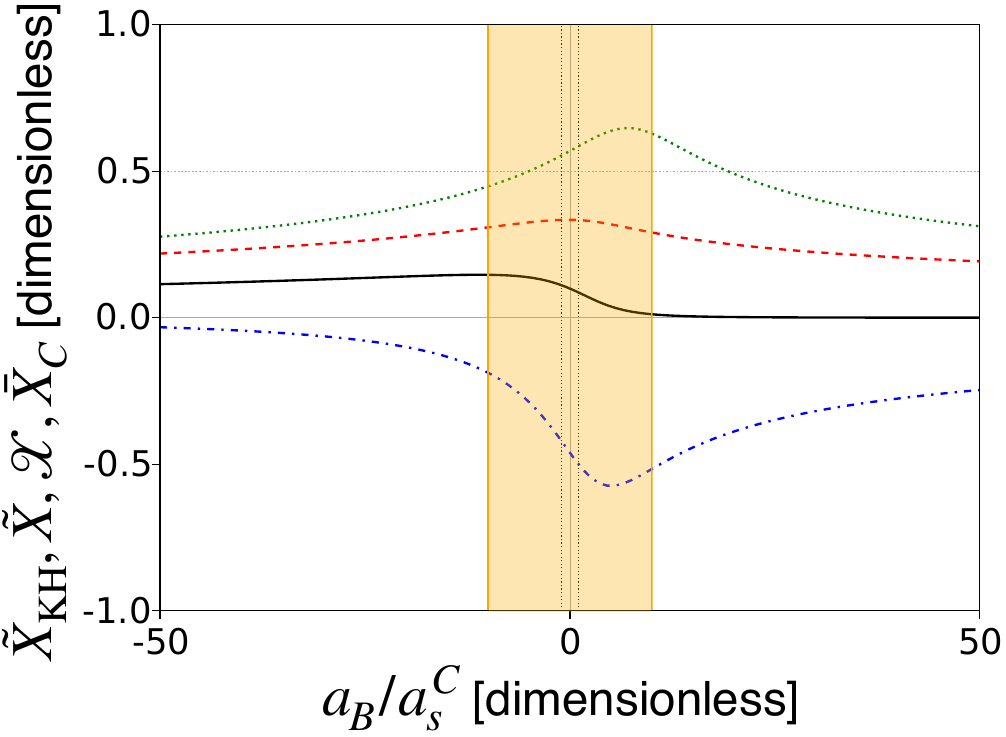}
\caption{Compositeness in the interpretation schemes introduced in Sec.~\ref{subsec:compositeness} as functions of the inverse scattering length $a_{B}/a_{s}^{C}$ in the attractive Coulomb plus short-range system. The solid, dashed, dash-dotted, and dotted lines represent $\tilde{X}_{\rm KH}$, $\tilde{X}$, $\mathcal{X}$, and $\bar{X}_{C}$, respectively.}
\label{fig:att-neg-X-vR}
\end{figure}

We then consider the compositeness $X$ of an eigenstate whose pole is located in the lower half of the $a_{B}k$ plane, where it evolves from a virtual state into a resonance in Fig.~\ref{fig:att-neg-large-k}. Since the compositeness of an unstable state is complex by definition, we employ the schemes introduced in Sec.~\ref{subsec:compositeness}. Figure~\ref{fig:att-neg-X-vR} shows $\tilde{X}_{\rm KH}$, $\tilde{X}$, $\mathcal{X}$, and $\bar{X}_{C}$ as functions of $a_{B}/a_{s}^{C}$ with a fixed effective range $r_{e}^{C}/a_{B} = -0.1$. We see that there are sizable differences among the schemes, implying substantial ambiguity in the interpretation of the complex compositeness. In particular, $\mathcal{X}$ (dash-dotted line) is negative, which cannot be interpreted as a probability. Here, for definiteness, we adopt $\bar{X}_{C}$ to interpret the internal structure of virtual states and resonances. Focusing on the region near $a_{B}/a_{s}^{C} = 0$, we find that $\bar{X}_{C}$ shows a peak structure. This may be due to the remnant of the short-range universality, which is expected to appear in the $- a_{B}/|r_{e}^{C}| < a_{B}/a_{s}^{C} < a_{B}/|r_{e}^{C}|$ region (shaded area). In the far-threshold region $a_{B}|k_{h}| \gg 1$ corresponding to $a_{B}/|a_{s}^{C}| \gg 1$, the compositeness approaches zero. This is in line with the analytic behavior of the compositeness discussed in Sec.~\ref{subsec:compositeness}.

As in the repulsive Coulomb case, we would like to examine the behavior of the compositeness when the Coulomb interaction is gradually turned off. However, as shown in Fig.~\ref{fig:att-neg-large-k}, the pole trajectory of the bound state is disconnected from that of the resonance while a bound state evolves to a resonance through a virtual state in the purely short-range system. This implies that it is not straightforward to trace the compositeness continuously to the no-Coulomb limit. It would be an interesting subject for future work to clarify how the disconnected trajectories are recovered in the $a_B \to \infty$ limit.


\section{Properties of eigenstates: positive effective range}
\label{sec:positive-re}

In the previous section, we considered a negative effective range $r_{e}^{C} < 0$, corresponding to $\sigma = +1$ in the EFT framework. However, as we will see in Sec.~\ref{sec:apply}, $r_{e}^{C}$ is positive in most physical systems. Therefore, we also examine a positive effective range, corresponding to $\sigma = -1$. The results for repulsive and attractive Coulomb interactions are presented in Secs.~\ref{subsec:rep-positive-re} and \ref{subsec:att-positive-re}, respectively.

When the effective range is positive, some care is required. As we will show below, a pole can appear in the upper half of the complex momentum plane with a finite real part, while such eigenmomenta are forbidden by the Hermiticity of the Hamiltonian. A similar behavior is known to occur in the absence of the Coulomb interaction~\cite{Ikeda:2011dx}. If such an unphysical pole exists, it originates from the neglect of higher-order terms in the scattering amplitude. These higher-order contributions become important for poles far from the threshold, and these poles exist outside the domain of validity of the truncated model. Since such poles are unphysical, from now on, we focus exclusively on the compositeness of states whose poles are located in the $a_{B}|k| < a_{B}/r_{e}^{C}$ region. 

\subsection{Repulsive Coulomb plus short-range interactions}
\label{subsec:rep-positive-re}

\begin{figure}
\centering
\includegraphics[width=0.45\textwidth]{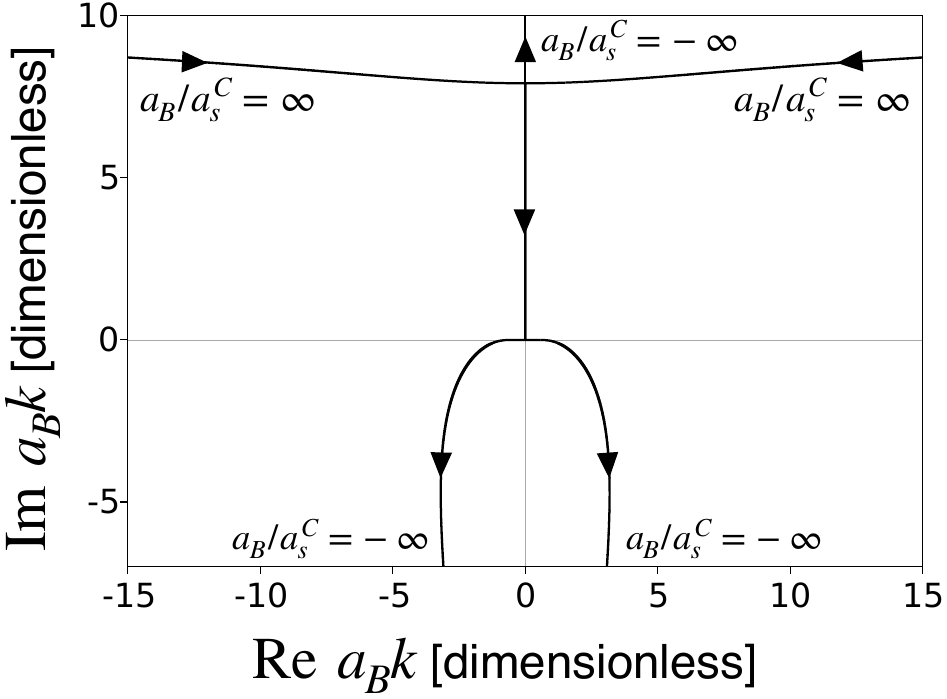}
\caption{Same as Fig.~\ref{fig:rep-neg-trajectory} but with the positive effective range $r_{e}^{C}/a_{B} = 0.1$.}
\label{fig:rep-pos-trajectory}
\end{figure}

Here, we consider the system in the presence of the repulsive Coulomb interaction with $r_{e}^{C} > 0$. First, we examine the pole trajectory in the complex momentum $a_{B}k$ plane for fixed effective range $r_{e}^{C}/a_{B} = 0.1$ in Fig.~\ref{fig:rep-pos-trajectory}. As in the case with the negative effective range, there are infinitely many virtual states originating from the Coulomb interaction near the origin, which is not shown in the figure. When the inverse scattering length $a_{B}/a_{s}^{C}$ is sufficiently large, there is a pair of poles with finite $\Re\,a_{B}k$ in the upper half plane. As $a_{B}/a_{s}^{C}$ decreases, these poles move towards the positive imaginary axis, and eventually collide at a point on the imaginary axis. This collision point is known as the exceptional point~\cite{Heiss:2012dx,Nishibuchi:2025uvt}. After that, one pole moves toward the threshold, while the other moves away from it. The former bound state turns into a resonance and an anti-resonance when $a_{B}/a_{s}^{C}$ becomes negative. This splitting of the bound state pole is also observed in the case with $r_{e}^{C} < 0$ in Fig.~\ref{fig:rep-neg-trajectory}, because in this near-threshold region, the pole behavior can be described by the zero-range theory~\cite{Mochizuki:2024dbf} in terms of the scattering length alone, irrespective of the effective range. As $a_{B}/a_{s}^{C}$ decreases further into the negative region, the imaginary part of the pole position increases in magnitude with an approximately constant real part. 

To discuss the domain of validity of this framework, we consider the position of the exceptional point. We numerically find that the exceptional point is located at $\Im\, a_{B}k \sim a_{B}/r_{e}^{C}$. This is in line with the expectation from the ERE for the purely short-range interaction, where the exceptional point exists exactly at $\Im\,k = 1/r_{e}$~\cite{Ikeda:2011dx,Hyodo:2013iga}. As $r_{e}^{C}/a_{B}$ is increased, the exceptional point moves closer to the origin and eventually enters the $a_{B}|k| < 1$ region. In this case, the domain of validity of the model becomes too narrow. Systems with such a large effective range should be treated with care within this model.

With sufficiently large $r_{e}^{C}/a_{B}$, the pole trajectory in the lower half of the $a_{B}k$ plane is qualitatively different from that in Fig.~\ref{fig:rep-pos-trajectory}. In this case, when the bound state pole crosses the threshold, it appears to be absorbed into the series of Coulomb-originated virtual state poles accumulated at the origin. Correspondingly, the $n$th virtual state in the limit $a_{B}/a_{s}^{C} \to \infty$ becomes the $(n-1)$th virtual state in the limit $a_{B}/a_{s}^{C} \to -\infty$. The $n = 1$ virtual state eventually becomes a deep virtual state in the limit $a_{B}/a_{s}^{C} \to -\infty$ as the virtual state originating mainly from the short-range interaction.

\begin{figure*}
\centering
\includegraphics[width=0.45\textwidth]{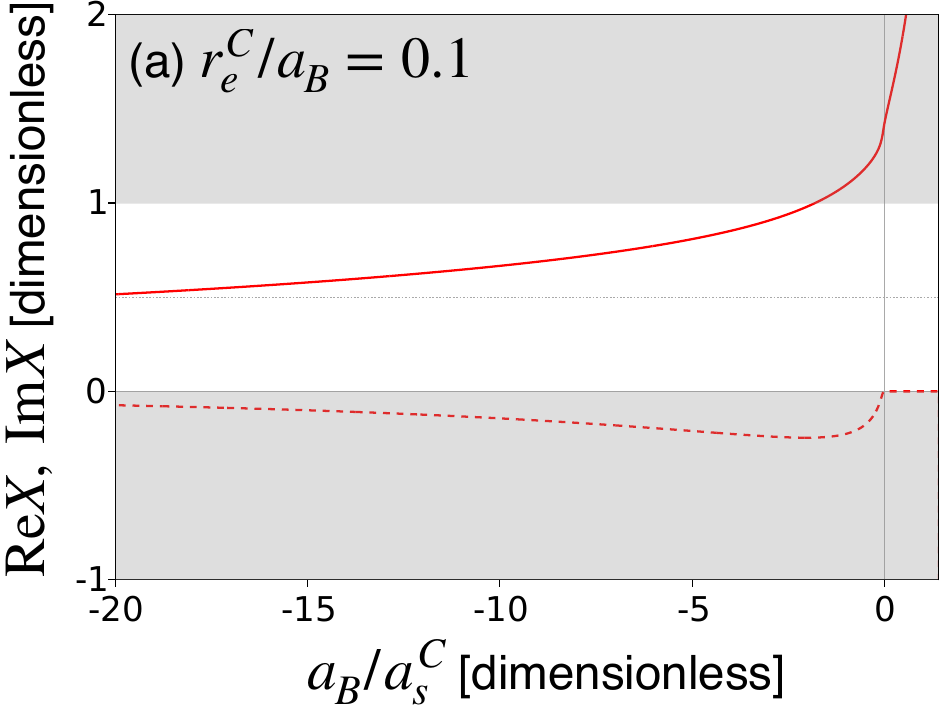}
\includegraphics[width=0.45\textwidth]{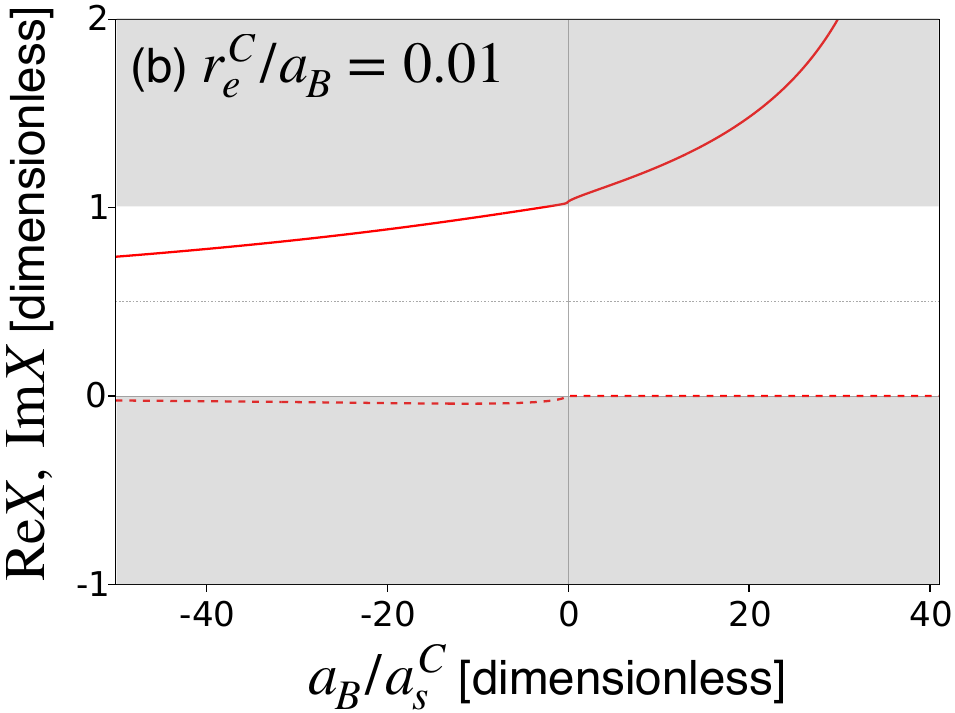}
\caption{Same as Fig.~\ref{fig:rep-neg-X} but with the positive effective range $r_{e}^{C}/a_{B} = 0.1$ [panel (a)] and $r_{e}^{C}/a_{B} = 0.01$ [panel (b)]. In the shaded region ($X > 1$ or $X < 0$), the compositeness $X$ cannot be regarded as a probability.}
\label{fig:rep-pos-X-r-0p1}
\end{figure*}

To quantitatively investigate the internal structure of eigenstates, we then evaluate their compositeness $X$. In Fig.~\ref{fig:rep-pos-X-r-0p1} we plot Re\, $X$ (solid lines) and Im\, $X$ (dashed lines) as functions of the inverse scattering length $a_{B}/a_{s}^{C}$. To observe the dependence on the effective range, we show the case with $r_{e}^{C}/a_{B} = 0.1$ [panel (a)] and $r_{e}^{C}/a_{B} = 0.01$ [panel (b)]. In the shaded region ($X > 1$ or $X < 0$), the compositeness $X$ cannot be regarded as a probability. We vary $a_{B}/a_{s}^{C}$ only within the range where no unphysical pole appears. 

In the $a_{s}^{C} > 0$ region, where the pole corresponds to a bound state, the compositeness exceeds unity ($X > 1.429$ and $X > 1.003$ for $r_{e}^{C}/a_{B} = 0.1$ and $r_{e}^{C}/a_{B} = 0.01$, respectively). This is because we chose $\sigma = -1$ to obtain a positive effective range within the truncated ERE as shown in Eq.~\eqref{eq:negative-Z}. This problem could be resolved by incorporating the higher-order contributions as a finite-range correction, thereby realizing a positive $r_{e}^{C}$ with $\sigma = +1$. An analogous situation is found in the deuteron, where the weak-binding relation in the zero-range limit gives $X \sim 1.7$~\cite{Kamiya:2017hni} due to the positive effective range, while the inclusion of finite-range corrections yields a reasonable value ($X < 1$)~\cite{Li:2021cue,Baru:2021ldu,Song:2022yvz,Albaladejo:2022sux,Kinugawa:2022fzn,Yin:2023wls}.

\begin{figure}
\centering
\includegraphics[width=0.45\textwidth]{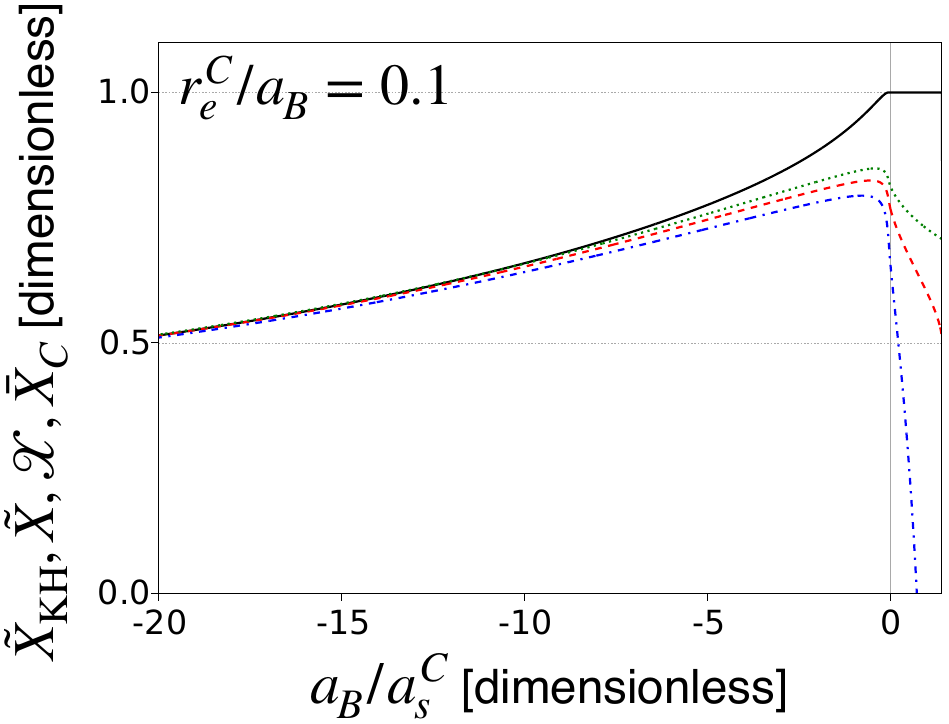}
\caption{Same as Fig.~\ref{fig:rep-neg-interpretations} but with the positive effective range $r_{e}^{C}/a_{B} = 0.1$. }
\label{fig:rep-pos-interpretations}
\end{figure}

Instead of incorporating a range correction, here we adopt the interpretation schemes for $X > 1$ to discuss the internal structure in the same manner in Sec.~\ref{sec:negative-re} for a negative effective range. In Fig.~\ref{fig:rep-pos-interpretations}, we show the results of the interpretations $\tilde{X}_{\rm KH}$ (solid line), $\tilde{X}$ (dashed line), $\mathcal{X}$ (dash-dotted line), and $\bar{X}_{C}$ (dotted line) as functions of the inverse scattering length. We first focus on bound states in the $a_{B}/a_{s}^{C} > 0$ region. In this region, the results differ quantitatively from each other. In particular, $\tilde{X}_{\rm KH}$ is $1$ by definition, as $X > 1$. On the other hand, $\tilde{X}$ and $\bar{X}_{C}$ are small. $\mathcal{X}$ is negative when $a_{B}/a_{s}^{C} \gtrsim 1$ for which a probabilistic interpretation is not possible. In this way, the difference among the results is not negligible for bound states. This indicates that the interpretation of the internal structure depends on the scheme. 

In the $a_{B}/a_{s}^{C} < 0$ region, states correspond to resonances and virtual states. The quantitative discrepancies among the interpretations become smaller as the pole moves away from the threshold. This is because, as shown in Fig.~\ref{fig:rep-pos-X-r-0p1}, $\Im\, X$ decreases as $a_{B}/a_{s}^{C}$ decreases, and $\Re\, X$ becomes smaller than unity at some point. When moving further away from the threshold, the state eventually becomes elementary dominant, with the compositeness becoming smaller than $0.5$. 

Taking the bound-state and resonance results in Fig.~\ref{fig:rep-pos-interpretations} together, the results near the threshold are all close to unity, although they show some quantitative variation. By comparing with panel (a) in Fig.~\ref{fig:rep-pos-X-r-0p1}, in panel (b), the deviation of $X$ from unity at the threshold is smaller, and the imaginary part of $X$ of resonances is also smaller. This suggests that for a smaller effective range $r_{e}^{C}/a_{B} = 0.01$, the near-threshold state is expected to become more composite. In this way, as in the negative $r_{e}^{C}$ case, a remnant of the short-range universality emerges in the weak Coulomb case.

\subsection{Attractive Coulomb plus short-range interactions}
\label{subsec:att-positive-re}

Finally, we discuss systems with an attractive Coulomb interaction and a positive effective range, $r_{e}^{C} > 0$. We begin by considering the pole trajectory in the complex momentum $a_{B}k$ plane with a fixed effective range. When $a_{B}/a_{s}^{C} \to \infty$, there are two unphysical poles in the upper half plane, symmetrically located about the imaginary axis, as in the repulsive Coulomb case with $r_{e}^{C} > 0$. As $a_{B}/a_{s}^{C}$ decreases, the two poles become bound-state poles with one moving toward the threshold and the other away from it along the imaginary axis. Among these two, the pole closer to the threshold eventually approaches the Coulomb ground-state level in the $a_{B}/a_{s}^{C} \to -\infty$ limit. This is the same behavior as in the case with $r_{e}^{C} < 0$.

\begin{figure}
\centering
\includegraphics[width=0.45\textwidth]{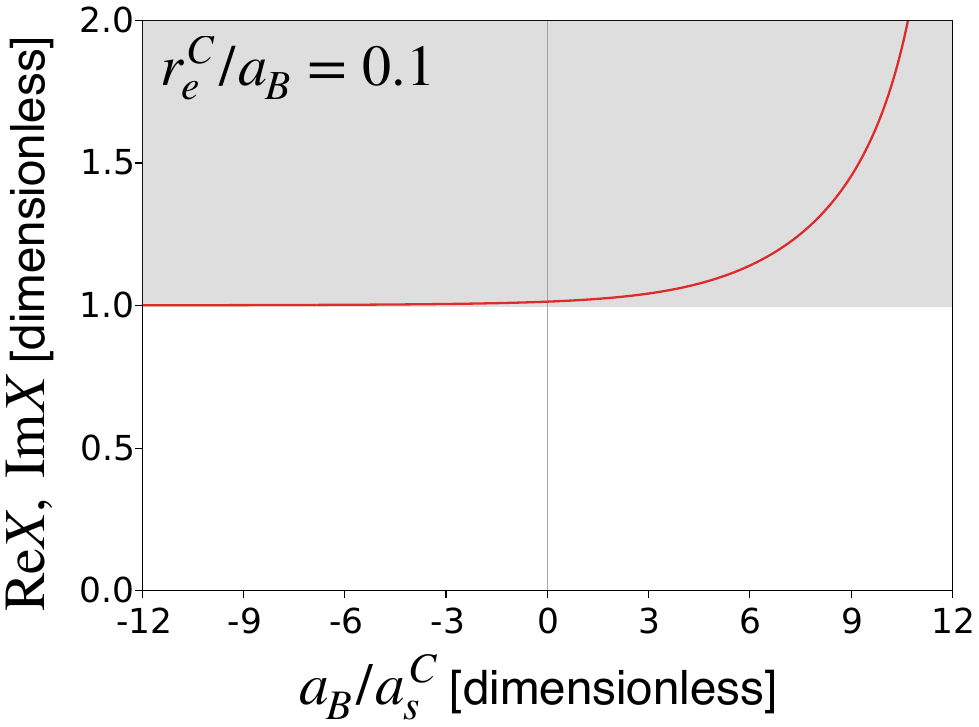}
\caption{Same as Fig.~\ref{fig:att-neg-X} but with the positive effective range $r_{e}^{C}/a_{B} = 0.1$. In the shaded region ($X > 1$ or $X < 0$), the compositeness $X$ cannot be regarded as a probability.}
\label{fig:att-pos-X}
\end{figure}

To investigate the internal structure of the eigenstates, we then compute the compositeness $X$ of the bound state. In Fig.~\ref{fig:att-pos-X}, we plot $X$ as a function of the inverse scattering length $a_{B}/a_{s}^{C}$ with $r_{e}^{C}/a_{B} = 0.1$. Figure~\ref{fig:att-pos-X} shows that as $a_{B}/a_{s}^{C}$ decreases, $X$ asymptotically approaches unity from above. In other words, $X$ remains larger than unity in the entire parameter region. This is in contrast to the $r_{e}^{C} < 0$ case, where $X$ approaches unity from below, as shown in Fig.~\ref{fig:att-neg-X}. 

\begin{figure}
\centering
\includegraphics[width=0.45\textwidth]{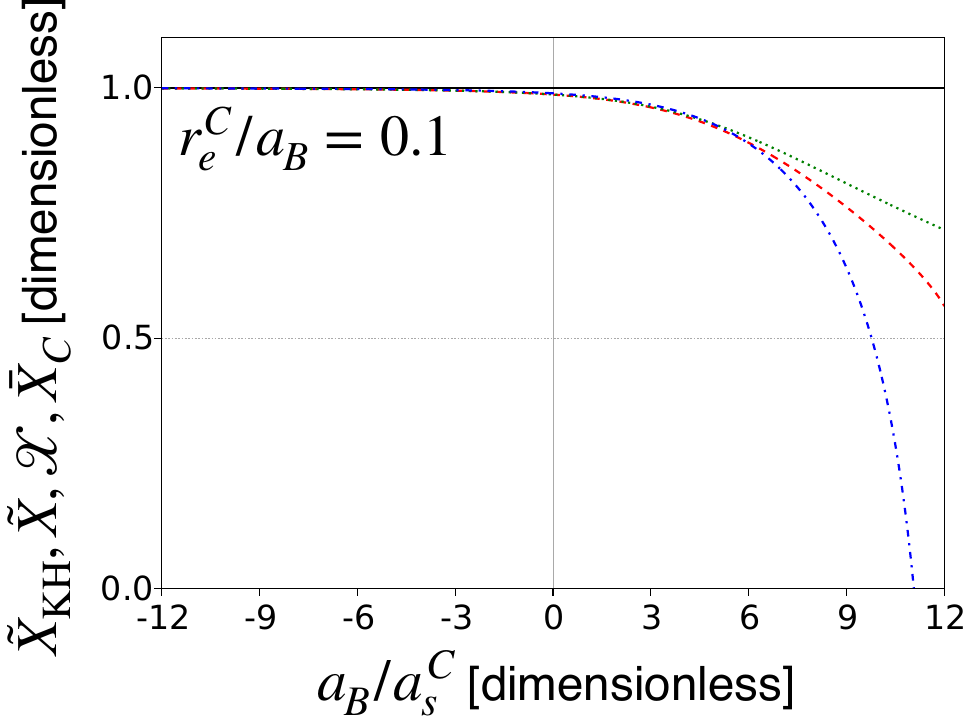}
\caption{Same as Fig.~\ref{fig:att-neg-X-vR} but with the positive effective range $r_{e}^{C}/a_{B} = 0.1$.}
\label{fig:att-pos-interpretations}
\end{figure}

To interpret the non-probabilistic compositeness $X > 1$, we evaluate $\tilde{X}_{\rm KH}$ (solid line), $\tilde{X}$ (dashed line), $\mathcal{X}$ (dash-dotted line), and $\bar{X}_{C}$ (dotted line) as functions of $a_{B}/a_{s}^{C}$ in Fig.~\ref{fig:att-pos-interpretations}. In the large $a_{B}/a_{s}^{C}$ region, the results differ among the schemes. This is because the deviation of $X$ from unity is large in this region, as shown in Fig.~\ref{fig:att-pos-X}. As $a_{B}/a_{s}^{C}$ decreases, all the results approach unity, with the ambiguity in the interpretation becoming smaller. This means that the bound state becomes increasingly composite dominant as it approaches the threshold.

\section{Application}
\label{sec:apply}

\subsection{Systems}
\label{subsec:systems}

\begin{table*}
 \caption{The properties of the scattering systems. $Z_{1}Z_{2}$ is the product of the electric charges, $m_{1}$ and $m_{2}$ are the masses, $\mu$ is the reduced mass, and $a_{B}$ is the Bohr radius. \label{tab:examples}}
 \begin{ruledtabular}
  \begin{tabular}{ccccccc}
    System & $Z_{1}Z_{2}$ & $m_{1}$ (MeV) & $m_{2}$ (MeV) & $\mu$ (MeV) & $a_{B}$ (fm) & Reference for $m_{1}$ and $m_{2}$\\ \hline
    $pp$ & $+1$ & 938 & 938 & 469 & 57.6 & \cite{ParticleDataGroup:2024cfk}\\
    $\alpha\alpha$ & $+4$ & 3727 & 3727 & 1864 & 3.63 & \cite{Mohr:2024kco}\\ 
    $\Omega^{-}\Omega^{-}$ & $+1$ & 1712 & 1712 & 856 & 31.6 & \cite{Gongyo:2017fjb}\\
    $\Omega_{ccc}^{++}\Omega_{ccc}^{++}$ & $+4$ & 4837 & 4837 & 2419 & 2.80 & \cite{Lyu:2021qsh}\\ 
    $\Xi^{-} \alpha$ & $-2$ & 1322 & 3727 & 976 & 13.9 & \cite{Mohr:2024kco,ParticleDataGroup:2024cfk}\\  
    $\Omega^{-} p$ & $-1$ & 955 & 1712 & 613 & 44.1 & \cite{HALQCD:2018qyu}\\ 
  \end{tabular}
  \end{ruledtabular}
\end{table*}

In this section, we apply the present formulation to evaluate the compositeness of near-threshold states in actual systems with Coulomb plus short-range interactions. The properties of the states considered here are summarized in Table~\ref{tab:examples}. $Z_{1}Z_{2}$ specifies the product of the electric charges of the particles. $m_{1,2}$ represents the masses of the particles, $\mu$ is the reduced mass, and $a_{B}$ denotes the Bohr radius. 

Let us first focus on the case of the Coulomb repulsion in the scattering of identical particles. As examples of scattering systems studied experimentally, we consider $pp$ and $\alpha\alpha$ scattering. The $s$-wave $pp$ scattering belongs to the ${}^{1}S_{0}$ channel. In the same channel, the $nn$ system is considered to possess a virtual state near threshold, whereas the corresponding structure in the $pp$ system is still not well understood. As a representative $\alpha$-cluster system, we consider the ${}^{8}$Be nucleus, which consists of two $\alpha$ particles, each with charge $Z_{i} = +2$. It is known that the two-$\alpha$ system would form a bound state if the Coulomb interaction were formally switched off, whereas in the physical case it appears as a resonance. In $pp$ scattering, the Bohr radius is much larger than the typical scale of the short-range interaction, which is of order 1~fm. This suggests that the Coulomb interaction does not strongly affect the short-range dynamics. On the other hand, the Bohr radius $a_B$ is smaller in the $\alpha\alpha$ system than in the $pp$ system because of the larger mass and charge of the $\alpha$ particle, and it becomes comparable to the short-range scale.

We then consider $\Omega^{-}\Omega^{-}$ and $\Omega_{ccc}^{++}\Omega_{ccc}^{++}$ scattering, for which scattering experiments have not yet been performed but near-threshold states have been suggested by lattice QCD calculations~\cite{Lyu:2021qsh,Gongyo:2017fjb}. The $\Omega^{-}$ is composed of three $s$ quarks and has an electric charge $Z = -1$. $\Omega_{ccc}^{++}$ consists of three $c$ quarks and carries charge $Z = +2$. This baryon has not yet been observed experimentally. Ref.~\cite{Gongyo:2017fjb} reports a shallow bound state in the $J = 0$ $\Omega^{-}\Omega^{-}$ channel. Because its Bohr radius is larger than the range of the short-range interaction, the Coulomb-originating eigenstates are well separated from the eigenstates mainly formed by the short-range interaction. The Bohr radius in the $\Omega_{ccc}^{++}\Omega_{ccc}^{++}$ system is comparable to the scale of the strong interaction, suggesting that the Coulomb interaction plays an important role. In fact, Ref.~\cite{Lyu:2021qsh} reports that there is a near-threshold resonance in the $J = 0$ $\Omega_{ccc}^{++}\Omega_{ccc}^{++}$ scattering, which would become a bound state if the Coulomb interaction were switched off. This indicates that the Coulomb interaction is essential in this system.
 
As examples of systems with an attractive Coulomb interaction, we investigate $\Xi^{-}\alpha$ and $\Omega^{-}p$ scattering, for which a bound state has not yet been observed experimentally but is theoretically expected. We consider $\Xi^{-}\alpha$ studied in Ref.~\cite{Hiyama:2022jqh}, based on the potential constructed from the HALQCD $\Xi N$ potential~\cite{HALQCD:2019wsz}. The $\Xi^{-}\alpha$ system is considered to be a Coulomb-assisted bound state, where the short-range attraction alone is insufficient to generate a bound state, but the additional Coulomb attraction leads to a bound state~\cite{Hiyama:2022jqh,Kamiya:2024diw}. The $\Omega^{-}p$ scattering in the $J = 2$ channel is studied by lattice QCD~\cite{HALQCD:2018qyu}. According to that study, the short-range $\Omega^{-}p$ interaction is attractive enough to form a bound state, while the attractive Coulomb interaction further enhances the binding energy~\cite{HALQCD:2018qyu}.

In Table~\ref{tab:examples}, we summarize the masses of the particles in the scattering systems considered here. The masses of $p$ and $\Xi$ are taken from the PDG~\cite{ParticleDataGroup:2024cfk}, and that of the $\alpha$ particle from Ref.~\cite{Mohr:2024kco}. For each scattering system studied in lattice QCD studies, we use the masses adopted in the corresponding lattice QCD calculation~\cite{Gongyo:2017fjb,Lyu:2021qsh,HALQCD:2018qyu} instead of the physical ones.

\subsection{Scattering parameters}
\label{subsec:scattering-parameters}

\begin{table}
 \caption{The Coulomb scattering length $a_{s}^{C}$, the Coulomb effective range $r_{e}^{C}$, and the ratio $|r_{e}^{C}|/a_{B}$.\label{tab:as-re}}
 \begin{ruledtabular}
  \begin{tabular}{ccccccc}
    System & $a_{s}^{C}$ (fm) & $r_{e}^{C}$ (fm) & $|r_{e}^{C}|/a_{B}$ & Reference\\ \hline 
    $pp$ & $-7.82$ & 2.83 & $0.049$ & \cite{Kong:1998sx} \\
    $\alpha\alpha$ & $-1.80 \times 10^{3}$ & 1.083 & $0.30$ & \cite{Higa:2008dn}\\ 
    $\Omega^{-}\Omega^{-}$ & $12.93$ & $1.21$ & $0.038$ & \cite{byYanLyu-san}\\
    $\Omega_{ccc}^{++}\Omega_{ccc}^{++}$ & $-19$ & $0.45$ & $0.16$ &\cite{Lyu:2021qsh}\\
    $\Xi^{-} \alpha$ & $-12.6$ & $6.66$ & $0.48$ & \cite{Kamiya:2024diw,byKamiya-san}\\
    $\Omega^{-} p$ & $3.2$ & $1.2$ & $0.029$ & \cite{byYanLyu-san}\\
  \end{tabular}
  \end{ruledtabular}
\end{table}

For the calculation of the compositeness in Eq.~\eqref{eq:wbr}, we need the Coulomb scattering length $a_{s}^{C}$ and the Coulomb effective range $r_{e}^{C}$. We summarize the values of $a_{s}^{C}$ and $r_{e}^{C}$ used in this work in Table~\ref{tab:as-re}. To observe the competition between the Coulomb and short-range interactions, $|r_{e}^{C}|/a_{B}$ is also listed. For the $pp$ and $\alpha \alpha$ scattering, $a_{s}^{C}$ and $r_{e}^{C}$ are taken from the EFT analyses in Ref.~\cite{Kong:1998sx} and Ref.~\cite{Higa:2008dn}, respectively. For the $\Omega^{-}\Omega^{-}$, $\Omega_{ccc}^{++}\Omega_{ccc}^{++}$, and $\Omega^{-}p$ scattering, we take the results obtained from the lattice QCD potentials combined with the Coulomb interaction~\cite{byYanLyu-san,Lyu:2021qsh}. For the $\Xi^{-}\alpha$ system, the results are based on the folding potential~\cite{Hiyama:2022jqh,Kamiya:2024diw,byKamiya-san}. 

As shown in this table, the sign of the scattering length differs depending on the system, while the effective range is positive in all cases. For systems with the repulsive Coulomb interaction, the absolute value of the scattering length is large; namely, the system is close to the unitary limit. Among those, $a_{s}^{C}$ for $\alpha\alpha$ scattering is extremely large. Focusing on the ratio $|r_{e}^{C}|/a_{B}$, we find that it is smaller than unity for all the systems considered, indicating that they belong to the weak-Coulomb regime, where remnants of short-range universality are expected to appear. In particular, this ratio is very small for $pp$, $\Omega^{-}\Omega^{-}$, and $\Omega^{-}p$ scattering.

\subsection{Eigenmomenta and eigenenergies}
\label{subsec:eigen}

\begin{table*}
 \caption{Eigenmomenta $k_{h}$ and eigenenergies $E_{h}$ obtained from the pole condition in Eq.~\eqref{eq:pole-condition-ERE}. Empirical $E_{h}$ is the eigenenergy obtained by analysis of experimental data or by theoretical predictions. The last column shows the property of each eigenstate classified based on Table~\ref{tab:poles}.\label{tab:E-k}}
 \begin{ruledtabular}
  \begin{tabular}{ccccc}
    Eigenstate & $k_{h}$ (MeV) & $E_{h}$ (MeV) & Empirical $E_{h}$ (MeV) & Property \\ \hline 
    $pp$ virtual state & $13 - 17i$ & $-0.14 - 0.47i$ & - & V \\
    ${}^{8}$Be & $18 - 8.8\times 10^{-5}i$ & $0.083 - 8.3\times 10^{-7}i$ & $0.092 - 3.4\times 10^{-6}i$~\cite{Rasche:1967urp} & R \\ 
    $\Omega^{-}\Omega^{-}$ dibaryon & $38i$ & $-0.85$ & $-0.7$~\cite{Gongyo:2017fjb} & B \\
    $\Omega_{ccc}^{++}\Omega_{ccc}^{++}$ dibaryon & $79 - 5.4i$ & $1.3 - 0.17i$ & - & R  \\
    $\Xi^{-} \alpha$ bound state & $21i$ & $-0.22$ & $-0.47$~\cite{byKamiya-san} & B \\
    $\Omega^{-} p$ dibaryon & $36i$ & $-1.07$ & $-2.46$~\cite{HALQCD:2018qyu} & B \\
  \end{tabular}
  \end{ruledtabular}
\end{table*}

By substituting $a_{s}^{C}$ and $r_{e}^{C}$ in Table~\ref{tab:as-re} into the pole condition in Eq.~\eqref{eq:pole-condition-ERE}, we calculate the eigenmomentum $k_{h}$ for each system. While Coulomb-originating eigenstates always exist near the threshold, we focus on the state mainly formed by the short-range interaction. For clarity in identifying the property of the states, the eigenenergy $E_{h} = k_{h}^{2}/(2\mu)$ is also shown. The eigenmomentum $k_{h}$ and eigenenergy $E_{h}$ calculated by $a_{s}^{C}$ and $r_{e}^{C}$ are summarized together with the empirical eigenenergy in Table~\ref{tab:E-k}. In the column ``Property,'' the state is classified as B for a bound state, V for a virtual state, and R for a resonance. 

In the $pp$ scattering, the energy of the eigenstate has a negative real part. Namely, the eigenstate corresponds to a virtual state. This is consistent with the existence of a virtual state in the neutral $nn$
system in the same $^{1}S_{0}$ channel. However, in the $pp$ scattering with the Coulomb interaction, the log cut prevents the existence of a virtual state on the imaginary momentum axis as discussed in Sec.~\ref{subsec:pole-condition}. Therefore, the $pp$ system is realized as a virtual state with a finite imaginary part of the eigenenergy. 

We then focus on the ${}^{8}$Be nucleus in the $\alpha\alpha$ scattering. Since the complex eigenenergy has a positive real part and a tiny negative imaginary part, it is identified as a narrow resonance. Our results are in good agreement with the resonance energy extracted from the phase-shift analysis of $\alpha\alpha$ scattering~\cite{Rasche:1967urp}. This indicates that the ${}^{8}$Be nucleus is indeed a near-threshold resonance for which the ERE provides a reasonable approximation. 

For the $\Omega^{-}\Omega^{-}$ dibaryon, the ERE yields a bound state with negative eigenenergy $E_{h} = -0.85$~MeV. This value reasonably agrees with the result $E_{h} = -0.7$~MeV obtained by solving the Schr\"odinger equation with the lattice QCD potential supplemented by the Coulomb interaction~\cite{Gongyo:2017fjb}. This demonstrates that the $\Omega^{-}\Omega^{-}$ bound state also lies within the region where the ERE is applicable. 

In contrast to the $\Omega^{-}\Omega^{-}$ dibaryon, the $\Omega_{ccc}^{++}\Omega_{ccc}^{++}$ dibaryon is concluded as an unbound state slightly above the threshold because of the negative scattering length~\cite{Lyu:2021qsh}. Based on the ERE, we find that the $\Omega_{ccc}^{++}\Omega_{ccc}^{++}$ dibaryon is a near-threshold resonance at $1.3 - 0.17i$~MeV. 

In Table~\ref{tab:E-k}, among the systems subject to the repulsive Coulomb interaction ($pp$, $^{8}$Be, the $\Omega^{-}\Omega^{-}$ dibaryon, and the $\Omega_{ccc}^{++}\Omega_{ccc}^{++}$ dibaryon), only the $\Omega^{-}\Omega^{-}$ dibaryon is a bound state, whereas the others are unbound. This reflects the fact that only the $\Omega^{-}\Omega^{-}$ scattering length is positive in Table~\ref{tab:as-re}. Furthermore, ${}^{8}$Be lies closest to the threshold, as expected from its large scattering length.

In the presence of an attractive Coulomb interaction, both $\Xi^{-}\alpha$ and $\Omega^{-}p$ are found to be bound states, as inferred from the values of $a_{s}^{C}$ and $r_{e}^{C}$ shown in Table~\ref{tab:as-re}. For $\Xi^{-}\alpha$, the binding energy obtained in previous work is $-0.47$~MeV~\cite{byKamiya-san}, while the ERE gives $-0.22$~MeV, providing a reasonable estimate of this value. For $\Omega^{-}p$, the ERE gives $E_{h} = -1.07$~MeV, which is in agreement with the lattice result, $E_{h} = -2.46$~MeV. These results indicate that, even in the presence of an attractive Coulomb interaction, the ERE provides a good description of the binding energy of shallow bound states.

While both $\Xi^{-}\alpha$ and $\Omega^{-}p$ are bound states, their scattering lengths have opposite signs. As discussed in Sec.~\ref{subsec:att-negative-re}, in the presence of the Coulomb attraction, a state remains bound regardless of variations in the scattering length. This explains why $\Xi^{-}\alpha$ is a bound state even though its scattering length is negative, unlike in the case with the purely short-range interaction.

\subsection{Compositeness}
\label{subsec:compositeness-apply}

\begin{table*}
 \caption{The compositeness $X$ of the eigenstates and those in the interpretation schemes introduced in Sec.~\ref{subsec:compositeness}.\label{tab:comp}}
 \begin{ruledtabular}
  \begin{tabular}{cccccccc}
    Eigenstate & $X$ & $\tilde{X}_{\rm KH}$ & $\tilde{X}$ & $\bar{X}_{C}$ & $\mathcal{X}$ & $\mathcal{Y}$ & $\mathcal{Z}$\\ \hline 
    $pp$ virtual state & $0.85 - 0.13i$ & $0.83$ & $0.81$ & $0.81$ & $0.80$ & $0.05$ & $0.15$ \\
    ${}^{8}$Be & $7.9 - 0.0013 i$ & $1.0$ & $0.53$ & $0.71$ & $-4.5$ & $11$ & $-5.5$\\ 
    $\Omega^{-}\Omega^{-}$ dibaryon & $1.5$ & $1.0$ & $0.76$ & $0.81$ & $0.64$ & $0.73$ & $-0.36$ \\
    $\Omega_{ccc}^{++}\Omega_{ccc}^{++}$ dibaryon & $1.5 - 0.2i$ & $0.99$ & $0.73$ & $0.79$ & $0.56$ & $0.86$ & $-0.42$ \\
    $\Xi^{-} \alpha$ bound state & $1.1$ & $1.0$ & $0.95$ & $1.0$ & $0.95$ & $0.09$ & $-0.04$ \\
    $\Omega^{-} p$ dibaryon & $1.2$ & $1.0$ & $0.86$ & $0.87$ & $0.84$ & $0.32$ & $-0.16$ \\
  \end{tabular}
  \end{ruledtabular}
\end{table*}

Finally, we discuss the internal structure of the eigenstates. We calculate the compositeness $X$ from $r_{e}^{C}$ and $k_{h}$ using Eq.~\eqref{eq:wbr}. Table~\ref{tab:comp} summarizes the compositeness $X$. By definition, the compositeness of unstable states ($pp$ virtual state, ${}^{8}$Be, and the $\Omega_{ccc}\Omega_{ccc}$ dibaryon) is complex. Furthermore, $X$ of the bound state (the $\Omega^{-} \Omega^{-}$ dibaryon, $\Xi^{-}\alpha$, and the $\Omega^{-}p$ dibaryon) is larger than unity in this formulation because of the positive effective range. For the probabilistic interpretation of these compositeness values, we calculate $\tilde{X}_{\rm KH}$, $\tilde{X}$, $\bar{X}_{C}$, and $\mathcal{X},\mathcal{Y},\mathcal{Z}$ introduced in Sec.~\ref{subsec:compositeness}. We note that when $\mathcal{X}$, $\mathcal{Y}$, or $\mathcal{Z}$ become negative, the structure of the state cannot be evaluated by the compositeness because of large uncertainty~\cite{Kinugawa:2024kwb}. 

For the $pp$ virtual state, all results show that it is composite dominant. The quantitative differences among the interpretation schemes are small, indicating that the uncertainty of the interpretation is small. This is explained by the fact that $0 < \Re\, X < 1$, while $\Im\, X$ is small.

${}^{8}$Be is composite dominant in most schemes, although the interpretation involves sizable ambiguity. The complex $X$ has a very small imaginary part, while its real part significantly exceeds unity. As seen in the case of a repulsive Coulomb interaction with a positive $r_{e}^{C}$ (Fig.~\ref{fig:rep-pos-X-r-0p1}), this model exhibits an enhancement of $\Re\, X$ at $a_{B}/a_{s} \sim a_{B}/r_{e}$. Since the $\alpha\alpha$ system has a large $|r_{e}^{C}|/a_{B}$, namely a small $a_{B}/|r_{e}^{C}|$, that point lies closer to threshold. Thus, $\Re\, X$ becomes large for ${}^{8}$Be, which lies very close to the threshold. Although most schemes support composite dominance, $\tilde{X}$ is evaluated as $0.53$, suggesting that the non-composite component cannot be neglected. This result reflects the fact that $\tilde{X}$ approaches $1/2$ for $|X| \gg 1$, and this tendency is observed for ${}^{8}$Be because of its large $\Re\, X$. In contrast, $\tilde{X}_{\rm KH} = 1$, since it approaches unity for $|X| \gg 1$. Thus, the different schemes lead to quantitatively different results. This interpretational ambiguity is also reflected in the large value of $\mathcal{Y}$, which characterizes the uncertainty of the interpretation. In nuclear physics, ${}^{8}$Be is known to possess an $\alpha$-$\alpha$ cluster structure, as established by experiments and few-body precise calculations~\cite{Wiringa:2000gb,Otsuka:2022bcf}. Although our results are subject to interpretational ambiguity, they indicate that the cluster component accounts for at least half of the ${}^{8}$Be wavefunction.

We now discuss the $\Omega^{-}\Omega^{-}$ and $\Omega^{++}_{ccc}\Omega^{++}_{ccc}$ dibaryons. In most cases, both dibaryons are composite dominant. For the $\Omega^{-}\Omega^{-}$ dibaryon, a lattice study of $\Omega^{-}\Omega^{-}$ scattering~\cite{Gongyo:2017fjb} shows that the mean distance between the two $\Omega^{-}$ baryons is about 4~fm, suggesting a spatially extended hadronic molecular structure. Our result is consistent with this picture. Compared with the result obtained without Coulomb interaction in Ref.~\cite{Kinugawa:2022fzn} ($0.79 \leq X \leq 1$), the present result is qualitatively consistent. This indicates that the Coulomb repulsion does not qualitatively modify the internal structure of the $\Omega^{-}\Omega^{-}$ dibaryon. With the $\mathcal{X}$, $\mathcal{Y}$, $\mathcal{Z}$ scheme, $\mathcal{Z} < 0$ and $\mathcal{Y}$ is sizable. This suggests that there is a substantial ambiguity in the interpretation of the internal structure. 

For the $\Omega_{ccc}^{++}\Omega_{ccc}^{++}$ dibaryon, in the absence of the Coulomb interaction, the weak-binding relation in Ref.~\cite{Kinugawa:2022fzn} gives $0.72 < X < 1$ for bound $\Omega_{ccc}^{++}\Omega_{ccc}^{++}$, which is qualitatively consistent with the present result. Although the Coulomb interaction changes the bound state to a resonance, its internal structure remains composite dominant. This is because the state still exists near the threshold, and the Coulomb interaction is weak with $|r_{e}^{C}|/a_{B} = 0.16$. Similar to the $\Omega^{-}\Omega^{-}$ case, $\mathcal{Z}$ is negative and $\mathcal{Y}$ is non-negligible in the $\mathcal{X}$, $\mathcal{Y}$, $\mathcal{Z}$ scheme, indicating a significant ambiguity for the interpretation.

For systems with an attractive Coulomb interaction, both $\Xi^{-}\alpha$ and $\Omega^{-} p$ are interpreted to be composite dominant. In Ref.~\cite{Kinugawa:2022fzn}, the compositeness of the $\Omega N$ dibaryon without Coulomb attraction is estimated to be in the range $0.8 \leq X \leq 1$. This quantitative agreement with the present analysis implies that the internal structure of $\Omega^{-} p$ is not significantly affected by the Coulomb interaction. Within the $\mathcal{X}$, $\mathcal{Y}$, $\mathcal{Z}$ scheme, the $\mathcal{Y}$ of $\Xi^{-}\alpha$ bound state is relatively small, indicating a small ambiguity. In this sense, $\mathcal{X}$ is consistent with the other results, while $\mathcal{Z}$ is only slightly negative.

In this way, we find an overall tendency toward composite dominance in all the states considered. At the same time, the degree of interpretational ambiguity depends on the state and on the prescription adopted. This result can be understood by the weak-Coulomb nature of the systems considered, and by the discussion in Secs.~\ref{sec:negative-re} and \ref{sec:positive-re}.


\section{Summary}
\label{sec:sum}

In this work, we have investigated near-threshold $s$-wave eigenstates in a two-body system with Coulomb plus short-range interactions. Motivated by the fact that the standard low-energy universality for short-range interactions is modified by the long-range Coulomb force, we formulate a nonrelativistic effective field theory and analyze the scattering amplitude, pole structure, and compositeness of the eigenstates.

We first present the full scattering amplitude by separating the pure-Coulomb contribution from the Coulomb-distorted short-range part. On this basis, we obtain the pole condition in the complex momentum plane and clarify the characteristic analytic structure generated by the logarithmic branch cut associated with the Coulomb interaction. We show that the Coulomb force qualitatively changes the threshold behavior of the poles. In particular, in the repulsive case, an $s$-wave bound state crosses the threshold and turns directly into a resonance, in contrast to the usual short-range case where a virtual state appears.

We then formulate the compositeness in terms of the energy derivative of the self-energy, which is suitable for the present system with the non-separable Coulomb interaction. For bound states, the resulting expression can be written solely in terms of the Coulomb scattering length $a_{s}^{C}$, the Coulomb effective range $r_{e}^{C}$, and the Bohr radius $a_{B}$. In this sense, we obtain a weak-binding relation for systems with Coulomb plus short-range interactions. This relation shows that the compositeness is governed by the interplay between the short-range scale and the Coulomb scale, and hence the usual threshold limit $X \to 1$ is generally modified by the presence of the Coulomb interaction.

By numerically examining both repulsive and attractive Coulomb interactions for negative and positive effective ranges, we find that the strength of the Coulomb interaction relative to the short-range interaction, characterized by $|r_{e}^{C}|/a_{B}$, plays a central role. When the Coulomb interaction is relatively weak, a remnant of the short-range universality survives, and near-threshold eigenstates tend to be composite dominant. In contrast, when the Coulomb interaction is strong, such enhancement of compositeness is absent, and even near-threshold states can become non-composite or elementary dominant. We also extend the discussion to resonances and compare several prescriptions for extracting interpretable compositeness from complex-valued quantities. In particular, when the remnant of the short-range universality persists with a weak repulsive Coulomb interaction, near-threshold resonances become composite dominant due to the continuity between bound states and resonances, leading to a qualitative difference from the case with only short-range interactions.

Finally, we apply the formalism to realistic systems, namely the $pp$ virtual state, $^{8}\mathrm{Be}$, the $\Omega^{-}\Omega^{-}$ dibaryon, the $\Omega_{ccc}^{++}\Omega_{ccc}^{++}$ dibaryon, the $\Xi^{-}\alpha$ bound state, and the $\Omega^{-}p$ dibaryon. We find an overall tendency toward composite dominance in these systems. At the same time, the degree of ambiguity in the interpretation depends on the state and on the prescription adopted, especially for resonances. These results indicate that the Coulomb interaction can significantly affect both the pole structure and the interpretation of the internal structure, while weak-Coulomb systems still retain characteristic features of composite near-threshold states.

The present study provides a unified framework for discussing the compositeness of near-threshold eigenstates beyond the short-range universality. It also offers a theoretical basis for understanding how Coulomb effects modify the threshold rule and the internal structure of weakly bound states and resonances in hadronic and nuclear systems.


\begin{acknowledgments}
The authors are grateful to Yan Lyu and Yuki Kamiya for useful discussions, in particular for providing the Coulomb scattering lengths and effective ranges for the $\Omega^{-}\Omega^{-}$, $\Omega^{-}p$, and $\Xi^{-}\alpha$ systems.
The authors thank 
Shunta Mochizuki,
Yusuke Nishida,
Christoph Hanhart,
Pascal Naidon
for fruitful comments and discussions.
This work was supported in part by the Grants-in-Aid for Scientific Research from JSPS (Grants
No. JP26K07088, 
No. JP26H01426, 
No. JP25K23387, 
No. JP23H05439, 
No. JP23KJ1796, 
No. JP22K03637). 
T.K. is supported by the RIKEN special postdoctoral researcher
program. 
\end{acknowledgments}

\bibliography{refs.bib}

\end{document}